\documentclass[useAMS,usenatbib]{mn2e}
\usepackage{graphicx}
\usepackage{txfonts}
\usepackage[T1]{fontenc}
\usepackage{aecompl}
\usepackage[usenames]{color}
\usepackage{enumerate}
\newcommand {\reac}[6] {$\rm\,{}^{#2}\kern-0.8pt{#1}\,({#3}\,,{#4})  \,{}^{#6}\kern-0.8pt{#5}\,$}
\newcommand{\object}{}

\newcommand{\Msun}{\mbox{${\rm~M_\odot}\,$}}

\newcommand{\HP}{\mbox{$H_{\rm{P}}\,$}}

\newcommand{\beq}{\begin{equation}}
\newcommand{\eeq}{\end{equation}}
\newcommand{\beqa}{\begin{eqnarray}}
\newcommand{\eeqa}{\end{eqnarray}}
\newcommand{\benu}{\begin{enumerate}}
\newcommand{\eenu}{\end{enumerate}}
\newcommand{\bite}{\begin{itemize}}
\newcommand{\eite}{\end{itemize}}
\newcommand{\bdes}{\begin{description}}
\newcommand{\edes}{\end{description}}

\newcommand{\comment}[1]{}
\title[Envelope Overshooting in Low Metallicity Intermediate- and High-mass Stars]{Envelope Overshooting in Low Metallicity Intermediate- and High-mass Stars: a test with the Sagittarius Dwarf Irregular Galaxy}
\author{}
\author[Tang et al.]{Jing Tang$^1$,  Alessandro Bressan$^1$, Alessandra Slemer$^2$,
Paola Marigo$^2$ \and Leo Girardi$^3$,  Luciana Bianchi$^4$, Phil Rosnfield$^2$,
 Yazan Momany$^3$\\
  $^1$ SISSA, via Bonomea 265, I-34136, Trieste, Italy,\\
  $^2$DIP FIS \&ASTRON,Vicolo dell'Osservatorio 4, I-37122 Padova, Italy \\
  $^3$OAPD, Vicolo dell'Osservatorio 5, I-37122 Padova, Italy \\
 $^4$ Dept of Physics and Astronomy, The Johns Hopkins University,
  407 Bloomberg center, 3400 N. Charles st., Baltimore, MD 21218, USA
}

\begin{document}
\date{Accepted 2015 October 22.  Received 2015 October 16; in original form 2015 August 20}

\pagerange{\pageref{firstpage}--\pageref{lastpage}} \pubyear{2015}

\maketitle

\label{firstpage}

\begin{abstract}
We check the performance of the {\sl\,PARSEC} tracks in reproducing the blue loops of
intermediate age and young stellar populations at very low metallicity.
We compute new evolutionary {\sl\,PARSEC} tracks of intermediate- and high-mass stars from 2\Msun to 350\Msun with enhanced envelope overshooting (EO), EO=2\HP and 4\HP, for very low metallicity, Z=0.0005. The input physics, including the mass-loss rate, has been described in {\sl\,PARSEC}~V1.2 version.
By comparing the synthetic color-magnitude diagrams (CMDs) obtained from the different sets of models with envelope overshooting EO=0.7\HP (the standard {\sl\,PARSEC} tracks), 2\HP and 4\HP, with deep observations of the Sagittarius dwarf irregular galaxy (SagDIG), we find
an overshooting scale EO=2\HP to best reproduce the observed loops. This result is consistent with that obtained by \citet{Tang_etal14} for Z in the range 0.001-0.004.
We also discuss the dependence of the blue loop extension
on the adopted instability criterion and find that, contrary to what stated in literature,
the Schwarzschild criterion, instead of the Ledoux criterion, favours the development of blue loops.
Other factors that could affect the CMD comparisons such as
differential internal extinction or the presence of binary systems
are found to have negligible effects on the results.
We thus confirm that, in presence of core overshooting during the H-burning phase,
a large envelope overshooting is needed to reproduce the main features of the central
He-burning phase of intermediate- and high-mass stars.
\end{abstract}


%%%%%%%%%%%%%%%%%%%%%%%%%%%%%%%%%%%%%%%%%%%%%%%%%%%%%%%%%%%%%%%%%%%%%%%%%%
\section{Introduction}\label{sec:introduction}
We have updated the new evolutionary tracks of massive stars from 14\Msun to 350\Msun in \citet{Tang_etal14} and thus built a complete library of evolutionary tracks from very low (M=0.1\Msun) to very massive (M=350\Msun) stars, from the pre-main sequence to the beginning of central carbon burning. These tracks are computed with {\sl\,PARSEC}: {\sl\,P}adova {T\sl{R}}ieste {\sl\,S}tellar
 {\sl\,E}volution {\sl\,C}ode. All the input physics, including the mass-loss rate, has been detailed in \citet{Bressan_etal12,Bressan_etal13}, \citet{Chen_etal14}, and \citet{Tang_etal14}. In this paper, we present the evolutionary tracks of intermediate- and high-mass stars at very low metallicity, Z=0.0005, and test them against the observations of the Sagittarius dwarf irregular galaxy (SagDIG).

\begin{figure*}
\includegraphics[angle=0,width=0.4\textwidth]{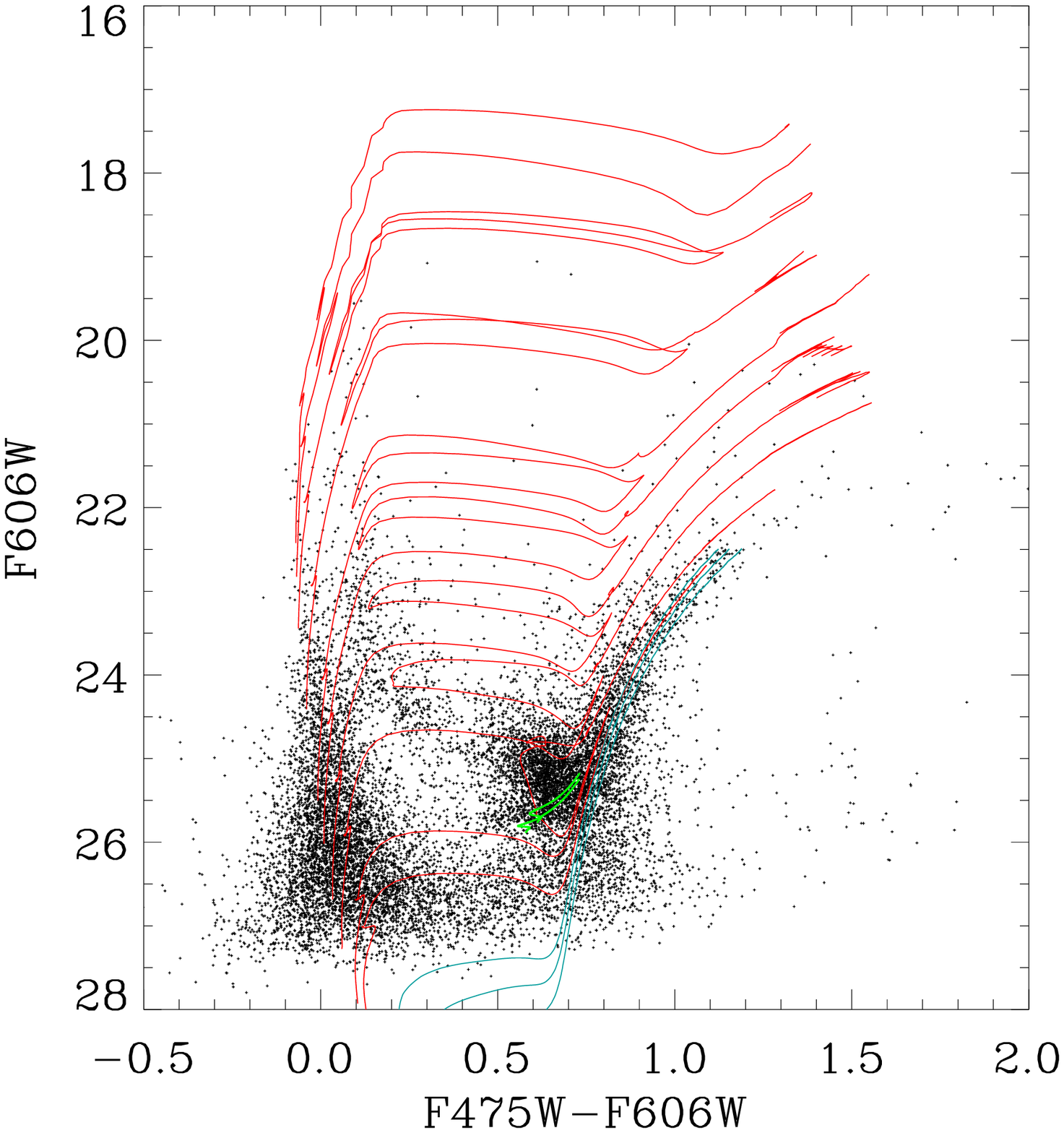}
\includegraphics[angle=0,width=0.4\textwidth]{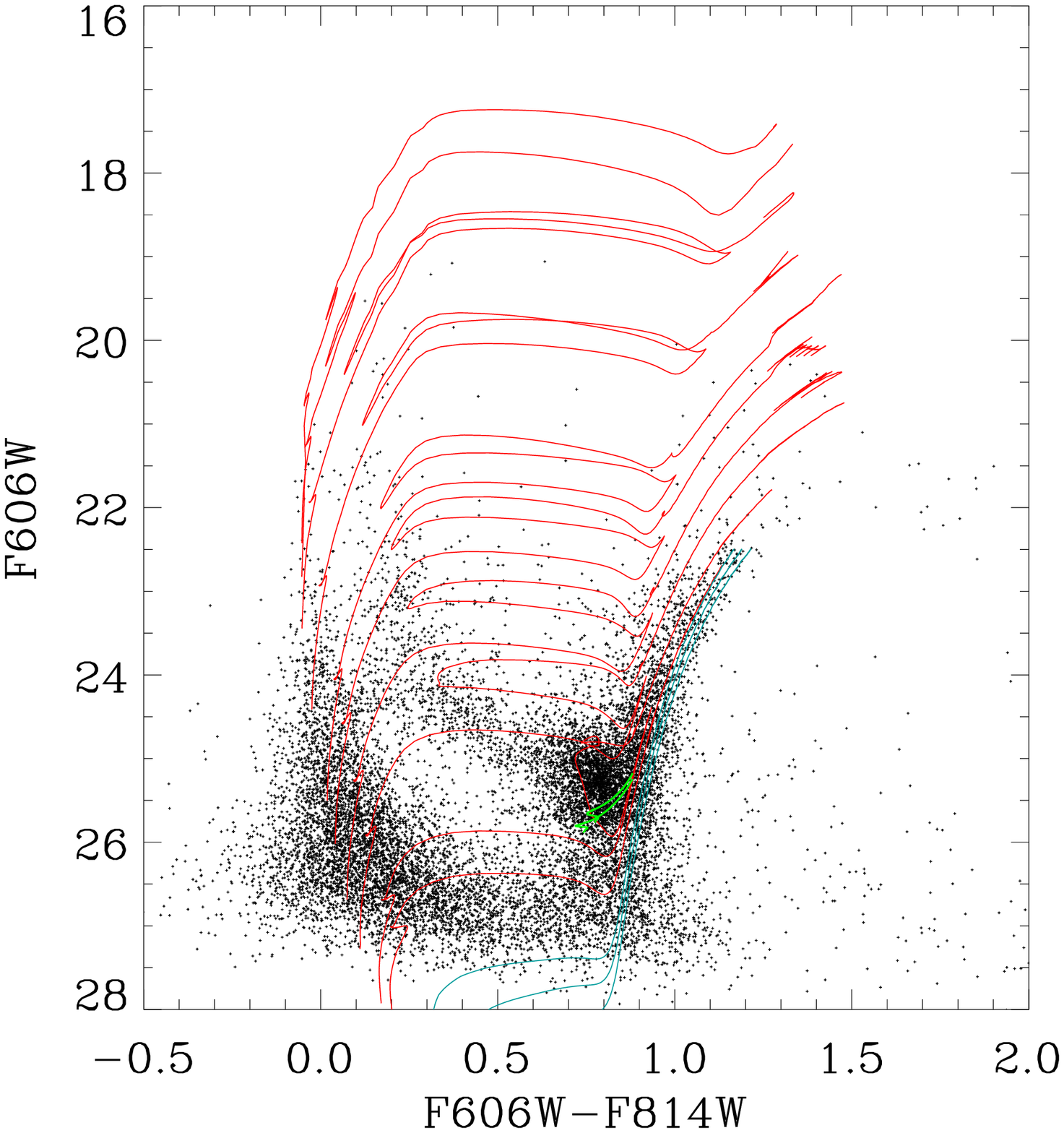}
\caption{Observed color-magnitude diagrams of SagDIG and evolutionary tracks computed by {\sl\,PARSEC}~V1.2 with EO=0.7\HP. Cyan lines are tracks of M=0.9\Msun, 1.0\Msun and 1.1\Msun with Z=0.0002, marking the RGB stars. Green lines mark HB evolutionary phases of stars with M=0.85\Msun and 0.95\Msun and with Z=0.0002. Red lines are tracks of M=1.5\Msun, 1.7\Msun, 2.3\Msun, 3.0\Msun, 4.0\Msun, 5.0\Msun, 8.0\Msun, 12.0\Msun, 16.0\Msun and 20.0\Msun with Z=0.0005. \label{fig_sagdig}}
\end{figure*}
Convection plays a crucial role in stellar structure and evolution, but it has not yet been fully understood. Convection is the macroscopic motions with energy and chemical element transport in the stellar interior. The most widely used theory for convection is the mixing length theory (MLT),
proposed by \citet{Bohm-Vitense58}, and the mixing length parameter is calibrated to be $\alpha_{MLT}$=1.74 based on the solar model \citep{Bressan_etal12}. In the MLT, the convective boundary is the location where the acceleration of the fluid elements is zero ($a$ = 0). However, at this border the elements still have a non-negligible velocity. They are able to cross the boundary and enter the stable radiative region until the velocity is zero ($v$ = 0). The region lies between $a$ = 0 and $v$ = 0 is called the overshooting. \citet{Bressan_etal81} provided a simple way to calculate the overshooting.
%two physical processes are still unclear up to now. One is the mass-loss rate, which plays a critical role over the lifetime of massive stars. The other is the internal mixing, from either differential rotation or convective overshooting.

As we know, convection occurs not only in the central core but also in the external envelope. Correspondingly, the extension of convective boundary is described by core overshooting above the convective core and envelope overshooting at the base of the convective envelope, respectively. Much theoretical and observational work supports the existence of core overshooting \citep{Bressan_etal81,Bertelli_etal85,Bertelli_etal90}. The main parameter describing core overshooting is the mean free path of convective elements across the border of the unstable region, $l_{c}=\Lambda_{c}$\HP, where \HP is the local pressure scale height. In this work $\Lambda_{c}$=0.5 is adopted \citep{Girardi_etal09}, which corresponds to 0.25\HP above the unstable region. The consideration of envelope overshooting in stellar models was first suggested by \citet{Alongi_etal91}. There are two relevant observational constraints: the location of RGB bump in globular clusters and old open cluster, and the extension of blue loops in intermediate- and high-mass stars. Both features can be reproduced better by considering a moderate amount of envelope overshooting EO=0.7\HP \citep{Alongi_etal91}.

However, \citet{Tang_etal14} found that a larger mixing scale, EO=2\HP or 4\HP, is preferred to reproduce the extension of the observed blue loops in nearby star forming dwarf galaxies. This result implies a strong dependence of the mixing scale below the formal Schwarzschild border of the envelope, on the stellar mass or luminosity. In this paper, we calibrate the envelope overshooting parameter by comparing the synthetic and observed color-magnitude diagrams (CMDs) of SagDIG, and further confirm this result.

The presence of blue loops during central He-burning phase was first thoroughly investigated by \citet{Lauterborn_etal71} who introduced the core potential $\phi=M_{core}/R_{core}$ to explain the occurrence. On the other hand, many studies show that the proximity of the H-burning shell to the H-He discontinuity marked by the depth of first dredge up triggers the blue loop \citep{Lauterborn_etal71,Robertson72,Stothers_etal91}. Thus any factor that moves the discontinuity deeper into the star causes a more extended loop. This conclusion is supported by \citet{Tang_etal14} who realized it by enhancing the envelope overshooting. Further \citet{Walmswell_etal15} demonstrated that this phenomenon is essentially due to the removal of the excess helium above the burning shell which causes changes to the mean molecular weight as well as the local opacity and the shell fuel supply. They believed that if this happens faster than the core evolution, the blue loop is triggered.

SagDIG is an ideal candidate to test our models at extremely low metallicity, because it is a very metal-poor star forming galaxy in the Local Group, and also it is nearby ($\approx$1.1 Mpc) which enables the {\em Hubble Space Telescope} to well resolve its star content. In section \ref{sec:data} we briefly describe the photometric data and the observed CMD from which the foreground contamination has been deducted. In section \ref{sec:models} we discuss the dependence of the blue loop extension on the adopted convection criterion, and present the new evolutionary tracks of intermediate- and high-mass stars computed by {\sl\,PARSEC} with different envelope overshooting. In section \ref{sec:scmd} we review the technique used to construct the synthetic CMDs, and in section \ref{sec:results} we show the comparison of different models with the observed CMD of SagDIG. Finally, we discuss and conclude in section \ref{sec:discon}.

\section{Data}\label{sec:data}
\subsection{Color-magnitude diagram}
To test our models, we use deep observations of SagDIG with the Advanced Camera for Surveys (ACS) on board the Hubble Space Telescope (HST). The observations contain two-epoch data-sets (GO-9820 and GO-10472) that are separated by $\sim$ 2 years. The main body of the galaxy ($\emph{l}=21.06^{\circ}$, $\emph{b}=-16.28^{\circ}$) was imaged in three filters: F475W, F606W and F814W. More details on the observations and data reduction can be found in \citet{Momany_etal14}.

Due to its low latitude, SagDIG suffers a heavy Galactic contamination but \citet{Momany_etal14} used two epochs data to analyse the relative proper motions of all detected stars and correct for foreground contamination. The CMDs cleaned for such a contamination of SagDIG are shown in Figure \ref{fig_sagdig},
for filters F475W and F606W  and for filters F606W and F814W, in the left and right panels,
respectively.

The high resolution HST/ACS observations enable us to distinguish various stellar populations in SagDIG,
as marked also by selected {\sl\,PARSEC} evolutionary tracks overplotted in the figures.
For the superposition of the evolutionary tracks we adopt a distance modulus (m-M)$_0$=25.06  and an extinction of
A(F475W)=0.657~mag, A(F606W)=0.520~mag and A(F814W)=0.286~mag, respectively. These values will be discussed
more extensively in a subsequent section.
SagDIG is characterized by the presence of very old populations. Red giant branch (RGB)
stars are indicatively marked by {\sl\,PARSEC} evolutionary tracks
of  masses M=0.9\Msun, 1.0\Msun and 1.1\Msun (cyan)
while, for horizontal branch (HB) stars we show two helium burning tracks of M=0.85\Msun and 0.95\Msun (pink).
Since these are meant to be the most metal poor stars in SagDIG,  we have adopted a very low metallicity Z=0.0002.
The turnoff stars of these old populations are not visible in the CMDs of Figure \ref{fig_sagdig}.
Indeed, the deeper region  of the observed main sequence,
around a magnitude of $m_{F606W}\sim$27, corresponds to the end of the H-burning phase of the track with
M$\sim$1.5\Msun, which is the faintest track plotted in red. This track is just near the
separation mass between low- and intermediate-mass stars, i.e. between those that undergo
or escape the helium flash. The track of M$\sim$1.7\Msun already belongs to the
intermediate-mass progeny and shows a well developed red clump.
For all the tracks with mass M$\geq$1.5\Msun, we use a larger metallicity, Z=0.0005, as will be discussed below.
We also note the presence of AGB stars in the continuation of the RGB tracks
above the corresponding RGB tips, and few of them likely corresponding to intermediate-mass stars. The AGB population is not modelled in our simulations because the majority of it comes from older stellar populations.

Starting from the red clump we may see the locus of the bluest He-burning  (BHeB) intermediate- and high-mass stars, which is clearly separated from the H-burning main sequence (MS).
This locus is marked by the blue loops of the other tracks plotted in the figure, which have
initial masses of M=2.3\Msun, 3.0\Msun, 4.0\Msun, 5.0\Msun,  8.0\Msun and 12.0\Msun.
We note that the upper main sequences in the SagDIG CMDs seem to extend beyond that
of the model of M=12\Msun and, for this reason, we also plot
the tracks with M=16.0\Msun and 20.0\Msun. However these tracks ignite He in the blue side of the CMDs
and do not perform blue loops as the less massive ones.

The clear separation of MS and BHeB stars makes the CMD of this galaxy,
especially the $m_{F606W}$ vs ($m_{F606W}-m_{F814W}$) CMD where the separation is striking, a powerful  workbench
for stellar evolution models at low metallicity.

\subsection{Metallicity of SagDIG}\label{sec:metallicity}
From the color of the RGB, \citet{Karachentsev_etal99} estimated the metallicity of SagDIG [Fe/H]=-2.45 $\pm$ 0.25, while \citet{Lee_etal00} derived [Fe/H] in the range from -2.8 to -2.4. By comparing the color differences between the RGB stars in the SagDIG and Galactic globular clusters (GGC) fiducial lines, \citet{Momany_etal02} yielded a mean metallicity [Fe/H]=-2.1 $\pm$ 0.2 for the red giants, and \citet{Momany_etal05} gave the range [Fe/H]=-2.2 to -1.9 depending on different assumed reddening.

On the other hand, by analysing optical spectrophotometry of $H_{\uppercase\expandafter{\romannumeral2}}$ regions in SagDIG, \citet{skillman_etal89} derived an oxygen abundance of 12+log(O/H)=7.42, which is in accordance with the measurement of \citet{Saviane_etal02} who estimated the O abundance in the range 12+log(O/H)=7.26 to 7.50.
%	O_H_sun=x_sun - 12  ;12+log10(O/H)=(grevesse)8.83 Z=0.017;   8.69 Z=0.0134  (Asplund)
From the latter values we obtain, using for the Sun 12+log10(O/H)=8.83 and Z=0.017 \citep{Grevesse_sauval98},
Z between 4.5E-4 and 7.9E-4. If instead we use the solar values  12+log10(O/H)=8.69 and Z=0.0134 \citep{Asplund_etal09},
we obtain Z between 5.0E-4 and 8.7E-4. We thus adopt for the young population of SagDIG a metallixity Z=5E-4 which is the lower value compatible with
spectroscopic observations.
We note that for the most recent populations  \citet{Momany_etal05} infer a metallicity between Z=0.0001 and Z=0.0004, by fitting
the extension of the blue loops of intermediate- and high-mass stars using the previous Padova isochrones.
\begin{figure}
\includegraphics[angle=0,width=0.4\textwidth]{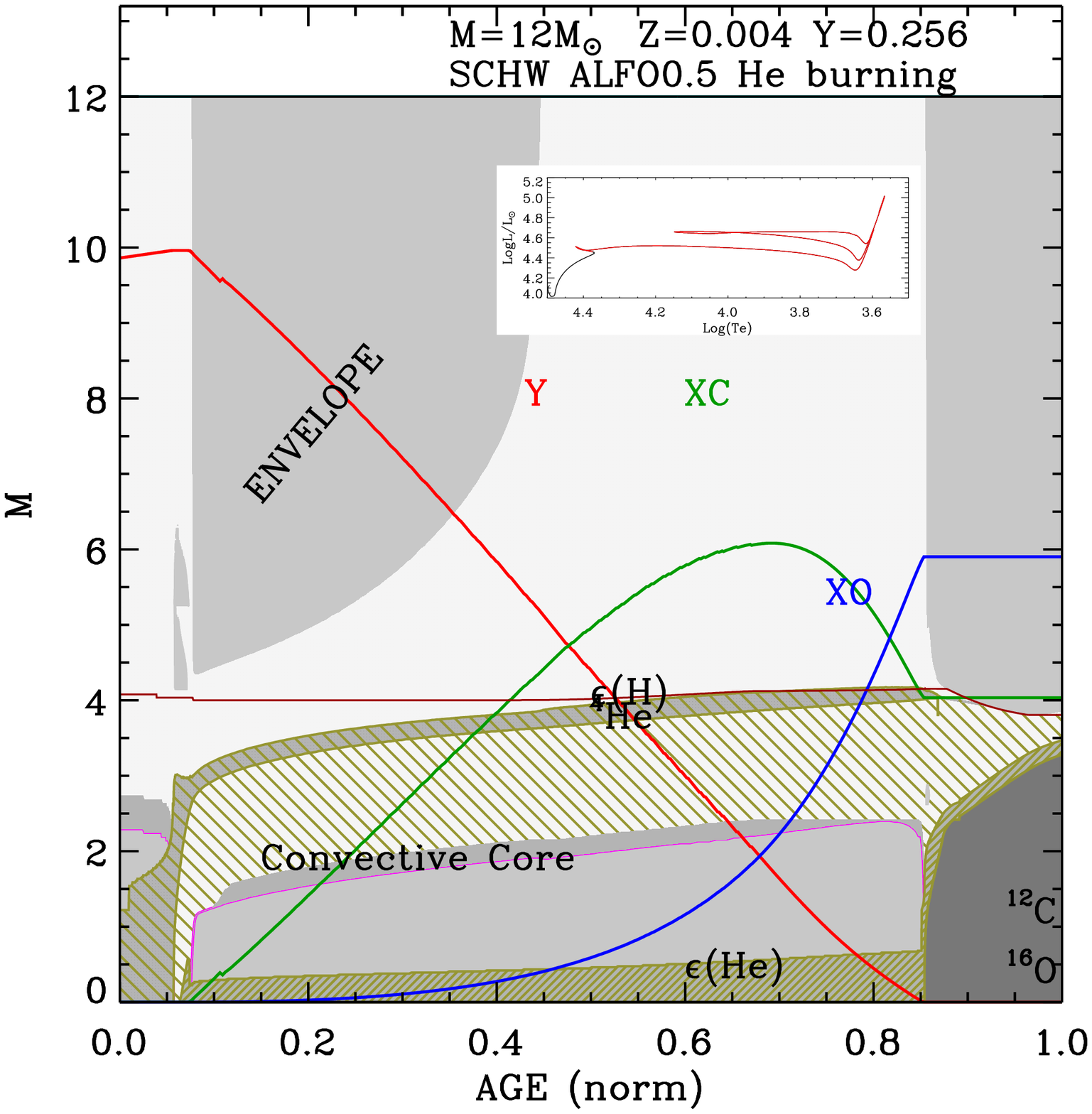}
\includegraphics[angle=0,width=0.4\textwidth]{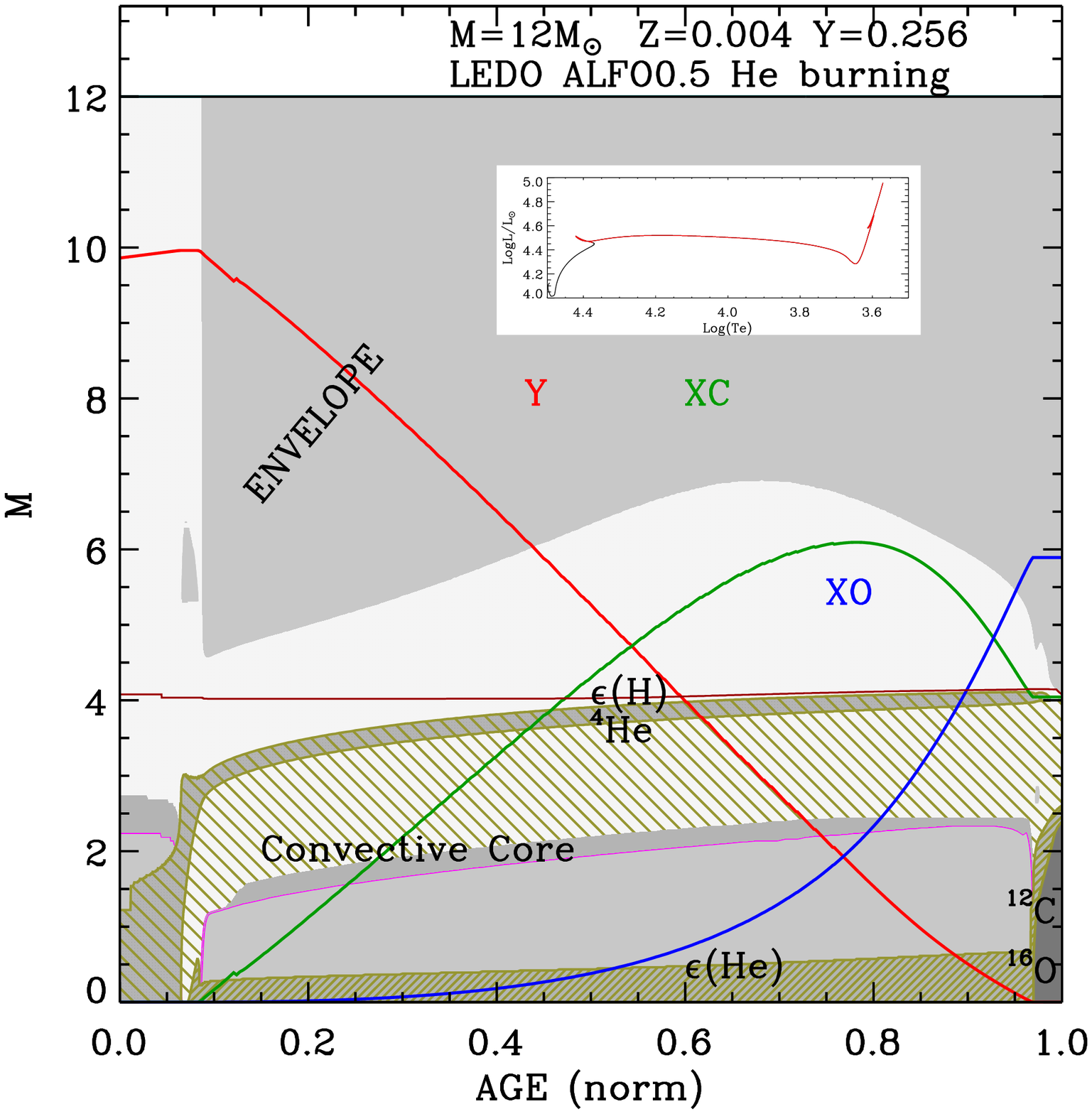}
\caption{Kippenhahn diagrams of central He-burning stars
 of $M$=12M$_\odot$ and $Z$=0.004 computed with different convection criteria,
the Schwarzschild criterion (left panel) and the Ledoux criterion (right panel). The core overshooting parameter $\Lambda_{c}$=0.5 is adopted. The brown line marks the location of the H-He discontinuity. The insets show the corresponding evolutionary tracks in the HR diagram. \label{convection_criteria}}
\end{figure}

\section{Models}\label{sec:models}
Not only is there a clear separation between the H-burning main sequence (MS) and the blue He-burning stars (BHeB) in the CMDs of Figure \ref{fig_sagdig},  but also red He-burning giants/supergiants (RSG) can be fairly well separated from the older red giant stars. This allows us to test different prescriptions used in building models of intermediate- and high-mass stars, in particular those that are known to affect the extension and duration of the blue loops. We have already performed a similar analysis on three well studied dwarf irregular galaxies (DIGs), \object{Sextans A}, \object{WLM} and \object{NGC 6822}, which are used to constrain
the models at  metallicities somewhat higher (Z=0.001-0.004) than the one suitable for SagDIG (Z=0.0005) \citep{Tang_etal14}. The result of this analysis indicated that, if one keeps the extent of the core overshooting parameter
fixed on the value resulting from comparison of low- and intermediate-mass stars (i.e. an overshooting length of about 0.25\HP above the unstable core), then the blue loops of intermediate- and high-mass stars are significantly reduced with respect to the models computed without core overshooting. To restore the extent of the blue loops, significant overshooting at the base of the convective envelope is required, with a typical extent of a few \HP.
It is worth noting that any mechanism that increases the size of the fully He exhausted region
tend to shorten or even suppress the blue loops.
This is also the case of models with rotational enhanced central mixing.
In some cases even models computed with the usual instability criterion and without any
extended mixing face the problem of lacking extended blue loops, especially at high metallicities.
The goal of this paper is to test if the conclusion obtained by \citet{Tang_etal14} remains valid also
at very low metallicities. For this purpose we model the brightest area of the CMD, which can be reasonably well represented by the last episode of star formation in SagDIG. We thus
focus on stars brighter than $m_{F606W}$=24, where the completeness is 100\%, 99.7\% and 98.8\% in F475W, F606W and F814W, respectively. This corresponds to a mass limit of $\sim$5.0\Msun and $\sim$3.0\Msun on the MS and on the He-burning phase respectively, as indicated in Figures \ref{fig_sagdig} and \ref{fig_sagdig_track_eo4}.

\subsection{Comparisons between the Schwarzschild criterion and the Ledoux criterion}
Before embarking in this comparison, we stress that {\sl\,PARSEC} models make use of the Schwarzschild criterion
for convective instability. The boundary of the convective region can be determined by either the Schwarzschild criterion or the Ledoux criterion.
According to Ledoux, the condition for the onset of convection is more restrictive than the Schwarzschild one
\begin{equation}
   \nabla_{rad} > \nabla_{ad} + \nabla_{\mu},
\end{equation}
where $\nabla_{rad}$, $\nabla_{ad}$ and $\nabla_{\mu}$ represent the radiative temperature gradient, adiabatic temperature gradient and molecular weight gradient, respectively. $\nabla_{\mu}$ can be expressed as
\begin{equation}
   \nabla_{\mu} = -(\frac{\partial ln{\rho}}{\partial lnT})_{P, \mu}^{-1}(\frac{\partial ln{\rho}}{\partial ln{\mu}})_{P, T}(\frac{dln{\mu}}{dlnP}).
\end{equation}
In a region with homogeneous chemical composition, $\nabla_{\mu} = 0$, as you'd expect, it becomes the Schwarzschild criterion,
\begin{equation}
   \nabla_{rad} > \nabla_{ad}.
\end{equation}
There have been claims in literature that the use of the Ledoux criterion could
eventually be more suitable in regions with variable mean molecular weight and it could give rise to more extended blue loops.
This has been shown in the pioneering paper by \citet{Chiosi_Summa70} who investigated the effect of the above two instability criteria on the morphology of the tracks and, specifically, on the extension of the blue loops. They found that the adoption of different criteria produces different size of semi-convective regions above the unstable core, and that only the model computed with the Ledoux criterion develops a blue loop.
Notably, the model computed with the Schwarzschild criterion ignites central He at high effective temperature, before reaching the red supergiant phase, and  only at the end of central He burning
it slowly moves toward the latter phase.
The different behaviour resulting from the adoption of different instability criteria may be understood by considering the competition between two important structural properties of the model,
the relative size of the H-exhausted core and the  mass pocket between the H-burning shell and the H-He discontinuity at the bottom of the convective envelope.
As already shown in \citet{Lauterborn_etal71},
since core overshooting during the H-burning phase
has the effect of increasing the relative mass size of the H-exhausted core, it favours He ignition and burning
in the red phase. On the other hand, it has been shown that the hydrostatic equilibrium location of He-burning models in the HRD is very sensitive to the mean molecular weight in the mass pocket between the H-exhausted core and the discontinuity in the H profile,
left either by the development of intermediate convective regions and/or by the penetration
of the convective envelope. \citet{Walmswell_etal15} have clearly shown that if the chemical
composition of this mass pocket,
which usually has an outward increasing H content, is artificially changed into a helium
rich mixture, the hydrostatic location of the model shifts to the blue region of the HRD.

This explains why the blue loop starts when the H-burning shell reaches the discontinuity of mean molecular weight
at the base of the H-rich envelope at early phase during the central He burning \citep{Tang_etal14}.
Three possibilities may arise, depending on when, eventually, the  H-shell reaches the H-discontinuity after central H-exhaustion. If it happens very soon during its expansion phase before the star becomes a red giant, the path in the HRD is inverted and the star ignites and burns He as a blue supergiant. Only at central He exhaustion the star will move toward the red giant phase. This  behaviour may be typical of most massive stars, especially at low metallicity. It corresponds to the case B (model computed with the Schwarzschild criterion) in \citet{Chiosi_Summa70}: that model does not perform a loop because it ignites He already in the blue loop hydrostatic configuration.
The second case happens when the H-shell reaches the discontinuity after He ignition in the red (super-)giant
phase, but early enough during central He burning. In this case the star performs a blue loop in the HRD as in case A (model computed with the Leodux criterion) of
\citet{Chiosi_Summa70}. Finally if the H-shell reaches the discontinuity at late time during central He burning, the star does not perform a blue loop and burns central He entirely as a red (super-)giant.
However, this behaviour is not a general property of the
adopted instability criterion. To demonstrate this we show here
two models with M=12\Msun computed with the two different instability criteria, and convective core overshooting is also considered in models. As shown in the Kippenhahn diagrams of Figure \ref{convection_criteria}, the internal structure of the models is identical at the end of central H burning
and both models are characterized by a rapid red-ward evolution followed by He ignition in the RSG phase,
independently of the adopted instability criterion. The larger size of the H-exhausted core originated
from the overshooting mixing forces both stars to ignite He in the supergiant phase.
%and accounting for convective core overshooting during the H-burning phase produces a relatively large H-exhausted cores and thus favours a rapid red-ward evolution  He ignition in the RSG phase, in both models.
However, in the case of the Schwarzschild criterion (upper panel), a large intermediate convective region develops, shifting the location of the H-He discontinuity slightly deeper, as indicated by the brown horizontal line, and effectively decreasing the size of the mass pocket between the H-exhausted core and the
H-He discontinuity.
Instead using the Ledoux criterion (lower panel), which is more restrictive, the intermediate convective regions are smaller or even suppressed, and the resultant mass pocket is relatively larger.
The difference with respect to the previous case is small, but in the former case
the H-burning shell is able to reach the discontinuity and a loop occurs, while in the latter case
this happens too late during He burning and the model does not perform a loop.
Thus the computations shown in Figure \ref{convection_criteria}  indicate  that, in presence of sizable
convective overshooting during central H-burning, the model computed with the Schwarzschild instability criterion
performs the blue loop while the one computed with the Ledoux criterion spends all the He-burning lifetime in the RSG phase.
We thus illustrate that the effect found by \citet{Chiosi_Summa70} is not a general property of the
instability criterion applied to massive stars. We instead confirm that the loop is activated if the H-shell
reaches the H-He discontinuity during the early He-burning phase \citep{Tang_etal14}. A possible explanation
of why this effect triggers a blue loop has been advanced by \citet{Walmswell_etal15}, who indicated that
this is the hydrostatic equilibrium location of a class of central He-burning models with a fully discontinuous H profile.

\subsection{Models with extended envelope overshooting}
\begin{figure}
\includegraphics[angle=0,width=0.4\textwidth]{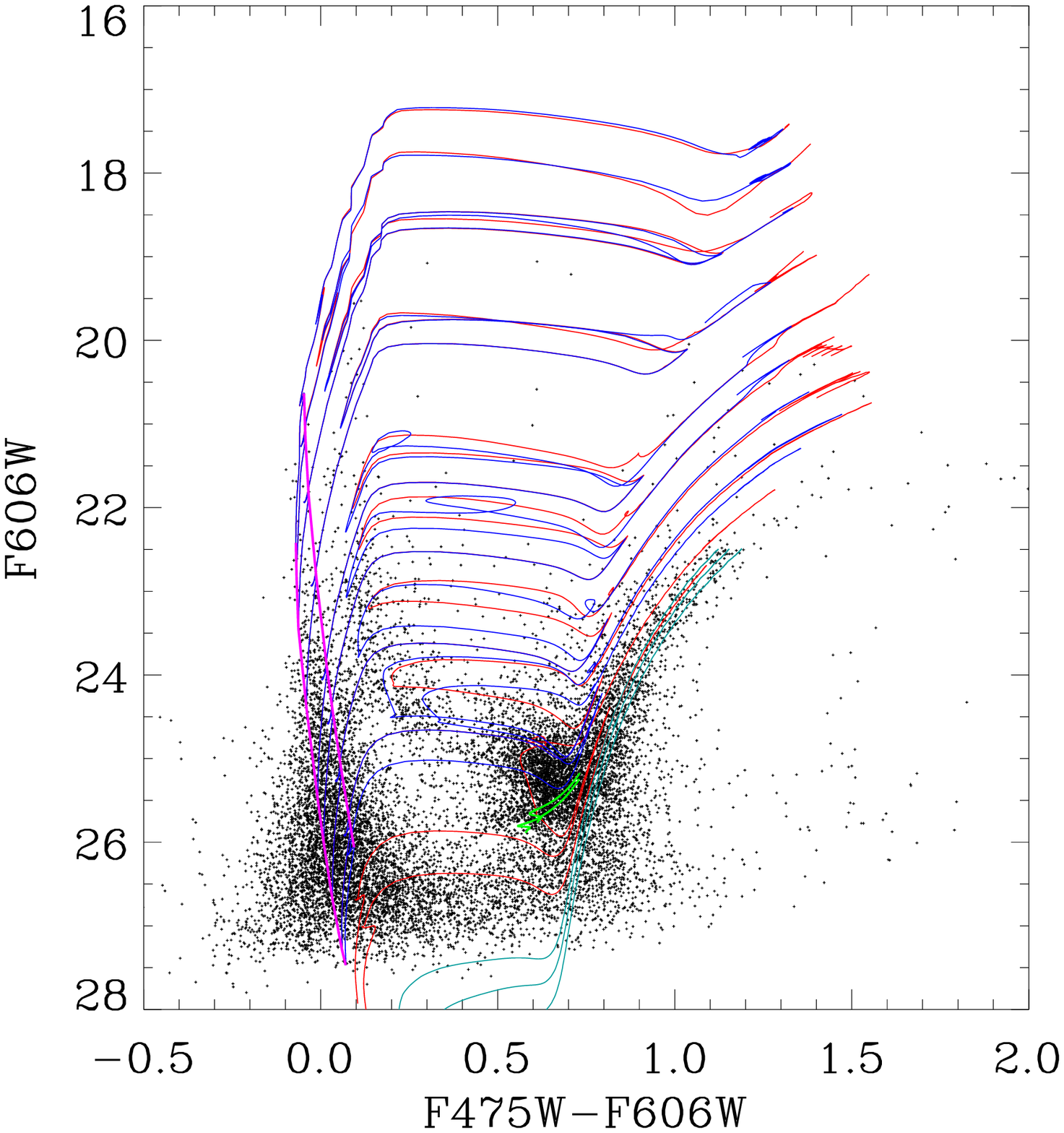}
\includegraphics[angle=0,width=0.4\textwidth]{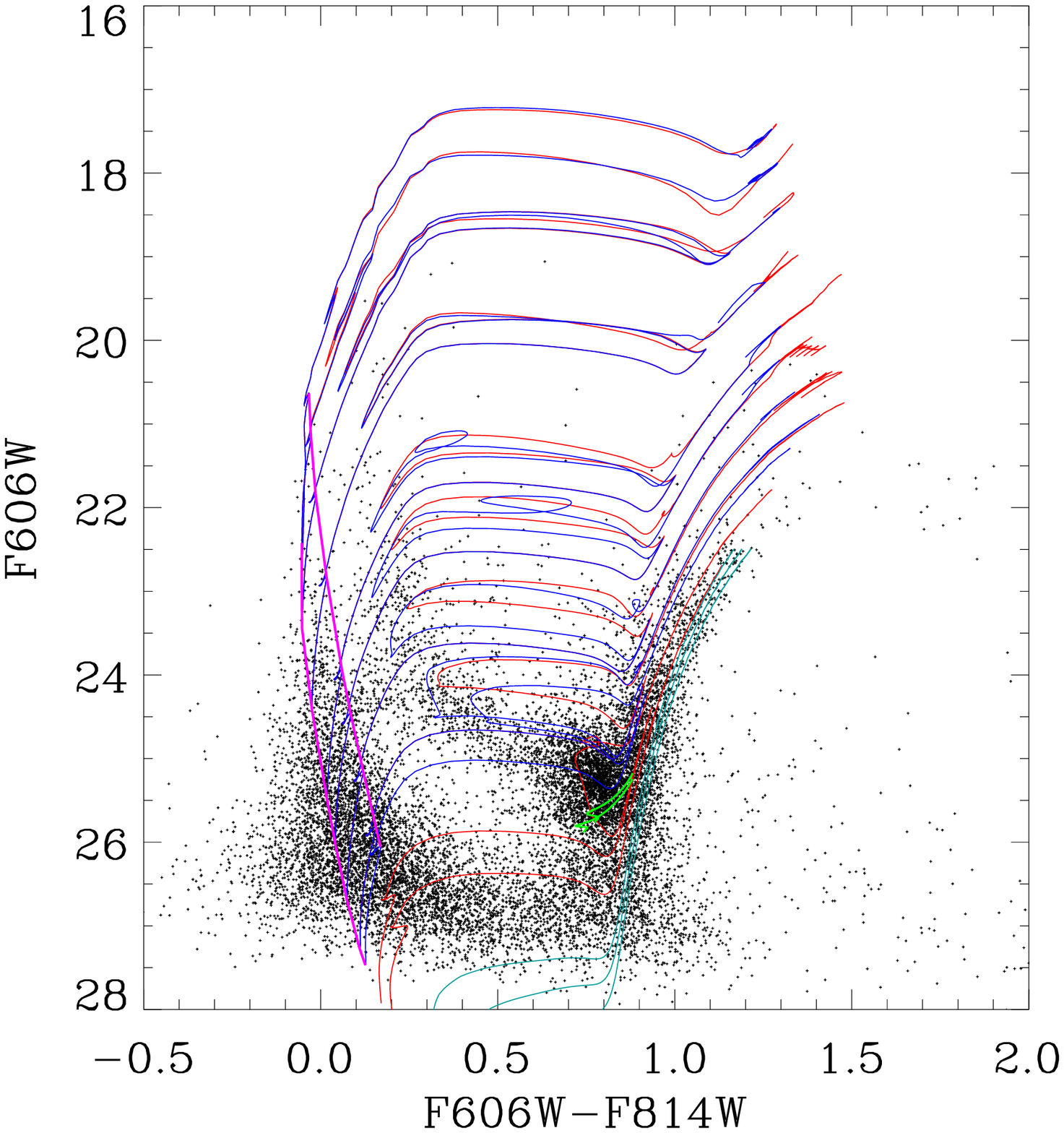}
\caption{Observed color-magnitude diagrams of SagDIG and evolutionary tracks computed by {\sl\,PARSEC}~V1.2 with EO=0.7\HP (red) and 4\HP (blue). Blue lines are tracks of M=2.1\Msun, 2.3\Msun, 3.0\Msun, 4.0\Msun, 5.0\Msun, 8.0\Msun, 12.0\Msun, 16.0\Msun and 20.0\Msun with Z=0.0005. Pink lines mark the beginning and end of central H burning of stars with different initial masses, respectively. \label{fig_sagdig_track_eo4}}
\end{figure}

To conclude this section we add to the CMDs of SagDIG in Figure \ref{fig_sagdig_track_eo4} the evolutionary tracks computed by assuming a large value of envelope overshoot, EO=4\HP, which is the largest value adopted in \citet{Tang_etal14}. These computations have been performed only for initial masses M$\geq$2.1\Msun and the
evolutionary tracks are shown in blue color.
The models
run superimposed to the standard {\sl\,PARSEC} models computed with EO=0.7\HP up to central He ignition.
Then, models with larger envelope overshooting ignite He at slightly lower luminosities
(the red giant tips are fainter) and they burn central He at a significantly lower luminosity,
both in the early red giant stage and in the blue loop phase.
An interesting effect of a large envelope overshooting is that,
in the low mass range, the new models cross the Cepheid instability strip, being 0.5 magnitude
fainter than the standard ones. We also note that the blue loop is significantly more extended than that in standard models.
These differences become smaller at increasing mass and practically disappear at masses
M$\geq$8\Msun.
As expected, the new models reproduce fairly  well the region of the observed blue He-burning stars of SagDIG.
We have also computed models with EO=2\HP, which lie between the two extreme cases already discussed.
In the next sections we compare the simulated CMDs obtained by these models with the observed one.

\section{Simulated color magnitude diagrams}\label{sec:scmd}
To construct a synthetic CMD of SagDIG, we follow the same procedure described in \citet{Tang_etal14}
as briefly summarized below.
For each of the three values of the envelope overshooting parameter, EO=0.7\HP, 2.0\HP and 4.0\HP, we generate a large mock catalogue in the theoretical plane with the following recipes.
\begin{enumerate}[1.]
\item The star formation rate (SFR) $\psi_m$ is specified as an exponential function of time, $\psi_m \propto exp(t/\tau)$, where $\tau$ is an adjustable parameter and $t$ is the stellar age, spanning from 1 Myr to 1 Gyr.
\item The age-metallicity relation (AMR) is replaced by a fixed value $Z$=0.0005 as we have discussed in section \ref{sec:metallicity}, since our analysis is limited to the youngest and brightest stars.
\item The initial mass function (IMF) $\phi_m$ is  a broken power law in the form of $\phi_m \propto m^{-\alpha}$,
where $\alpha = 0.4$ for $0.1 \Msun \leq m < 1 \Msun$, while the exponent is parametrized for $1 \Msun \leq m \leq M_{UP}$. The total mass is normalized to 1 in order to express SFR in units of \Msun/yr.
\item The library of theoretical isochrones is computed by {\sl\,PARSEC}, giving the luminosity $L$, the effective temperature $T_{eff}$, the surface gravity $g$ and other physical properties of stars as a function of stellar age $t$, initial mass $M_{ini}$, and metallicity $Z$.
\end{enumerate}
We transform the  catalogue  into the observational plane by means of
bolometric corrections, as a function of $T_{eff}$, $g$ and $Z$, adopted from \citet{Marigo_etal08}
and accounting for the distance. We finally add the effect of extinction and photometric errors. We may also include
the effect of binarity as discussed later.
With the simulated catalogue that contains much more stars than the observed data, we perform
hundred realizations of the CMD models to obtain, on one hand, the average and variance of the star luminosity function and, on the other hand, the best-fit model to the observed luminosity function.
We do not try more sophisticated statistical methods to reproduce the observed CMD, because
our aim here is to obtain the best value of the envelope overshooting parameter to be used
with the fixed metallicity Z=0.0005.
Since we care about the extension of the blue loops in intermediate- and high-mass stars, we just select stars with mass M$\geq$1.9\Msun, and also set the apparent magnitude limit to $m_{F606W}$=24 in the simulation.

\subsection{Simulated photometric errors}
\begin{figure}
\includegraphics[angle=0,width=0.4\textwidth]{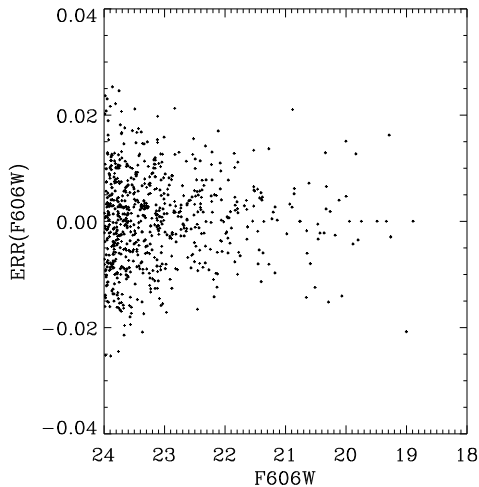}
\caption{Simulated photometric errors as a function of the apparent magnitude in the F606W filter.\label{fig_error}}
\end{figure}

\citet{Momany_etal14} have estimated the photometric errors for observed stars in SagDIG from artificial star experiments.
To account for photometric errors, we first bin their results in 0.1 mag steps
as a function of the apparent magnitude in each filter, and  calculate the median error for each bin.
Then we assign to each star in the mock catalogue an error that is randomly drawn from a Gaussian distribution with the standard deviation derived from the median corresponding to its magnitude \citep{Tang_etal14}.
We note that since  the observations are very deep, the simulated errors for stars brighter than 24 mag are small ($\leq$0.03~mag), as can be seen in Figure \ref{fig_error}.

\subsection{Foreground and internal extinction}
\citet{Lee_etal00} estimated a low foreground reddening E(B-V)=0.06 based on the (B-V) vs (V-I) diagram, consistent with the value E(B-V)=0.07 derived by \citet{Momany_etal02} from the (V-I) color distribution of foreground stars toward SagDIG, as the blue cut-off of the (V-I) location is a function of reddening along the line of sight. \citet{Demers_etal02} inferred E(B-V)$\approx$0.05 in the same way, but using (R-I) color distribution. The infrared dust maps of \citet{Schlegel_etal98} indicate a slightly higher reddening E(B-V)=0.12.
On the other hand, spectroscopic studies of $H_{\uppercase\expandafter{\romannumeral2}}$ regions in SagDIG suggest a higher reddening. \citet{skillman_etal89} calculated $c(H_\beta)$=0.33 and E(B-V)=0.22 is derived according to the relation $c(H_\beta)=1.47E(B-V)$ \citep{Seaton79}. \citet{Saviane_etal02} obtained a similar value E(B-V)=0.19 from the measurement of the Balmer decrement. As this method is based on the H line ratio, the estimated value include both  the foreground and the internal reddening.
As young stars are physically associated with the warm interstellar medium (ISM), it is reasonable to believe they suffer higher reddening compared to old stars. This trend was also found in other dwarf irregular galaxies \citep{Bianchi_etal12,Tang_etal14}.
\citet{Bianchi_etal12} derived individual star by star extinction from their multi-band data
of \object{Sextans A}, \object{WLM} and \object{NGC6822} dwarf irregular galaxies and this information was implemented in the
synthetic CMD analysis by \citet{Tang_etal14}.
Since for SagDIG we lack multi-band data and cannot repeat the same procedure with the same accuracy, we deal with the extinction in the following ways. The simplest way is to use a single value of the attenuation in each photometric band, as derived from the simultaneous alignment of the observed and modelled MS stars in both the $m_{F606W}$ vs ($m_{F475W}-m_{F606W}$) and $m_{F606W}$ vs ($m_{F606W}-m_{F814W}$) CMDs.

For the superposition of the evolutionary tracks, we adopt a distance modulus (m-M)$_0$=25.06 and an extinction of
A(F475W)=0.657~mag, A(F606W)=0.520~mag and A(F814W)=0.286~mag, respectively. These values will be discussed
more extensively in a subsequent section.

\section{Results}\label{sec:results}
\begin{figure*}
\includegraphics[angle=0,width=0.4\textwidth]{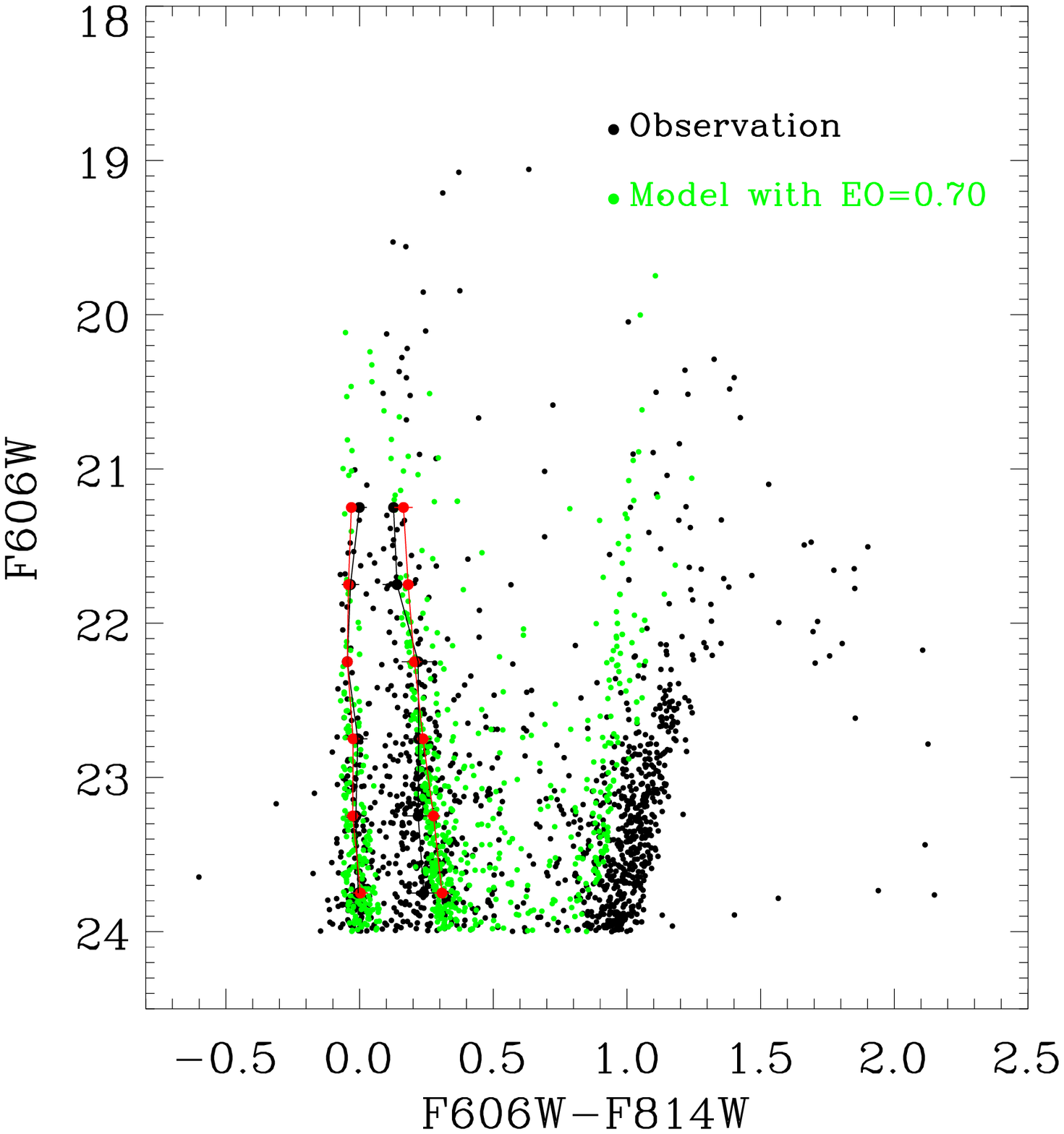}
\includegraphics[angle=0,width=0.4\textwidth]{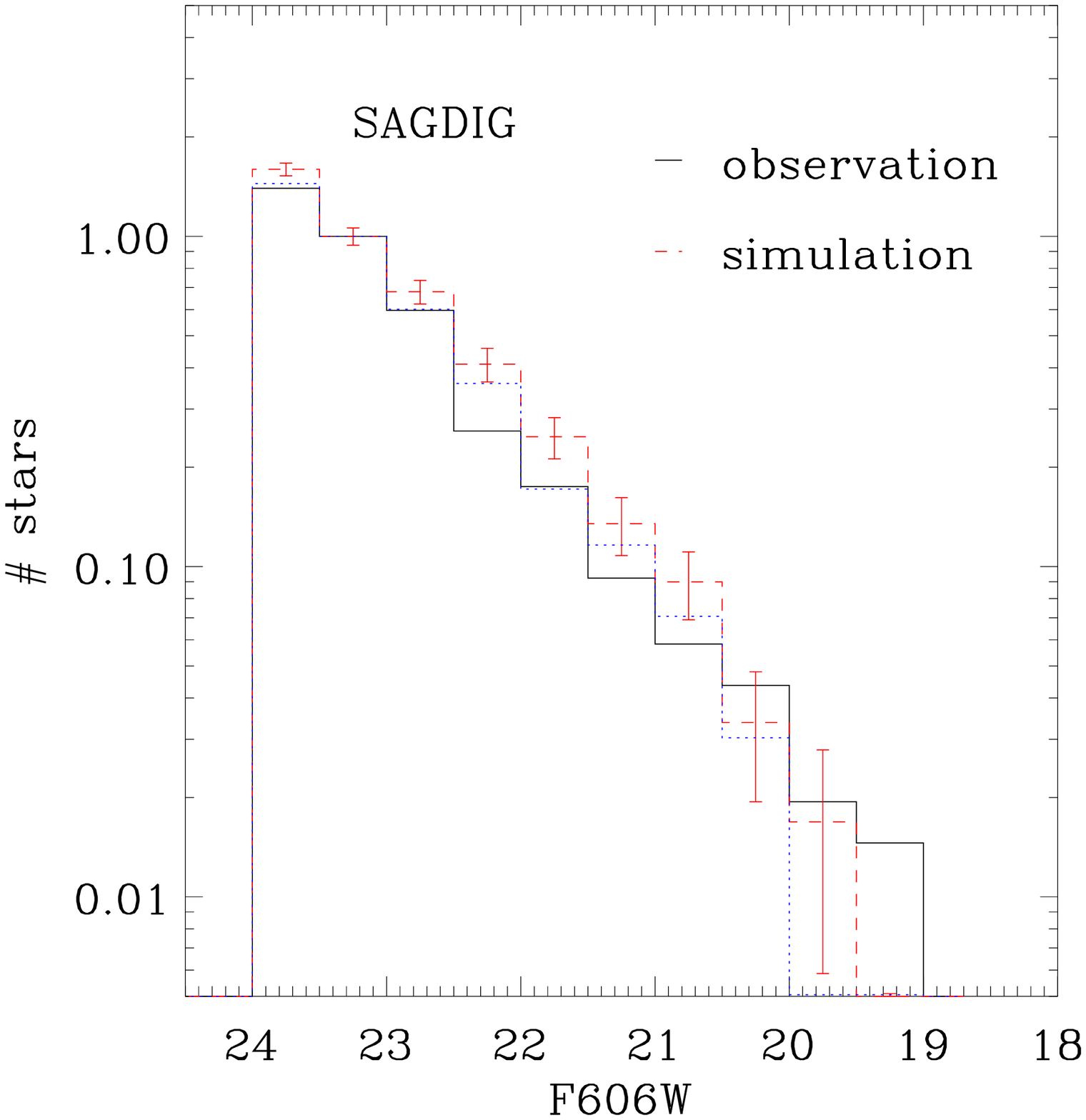}
\includegraphics[angle=0,width=0.4\textwidth]{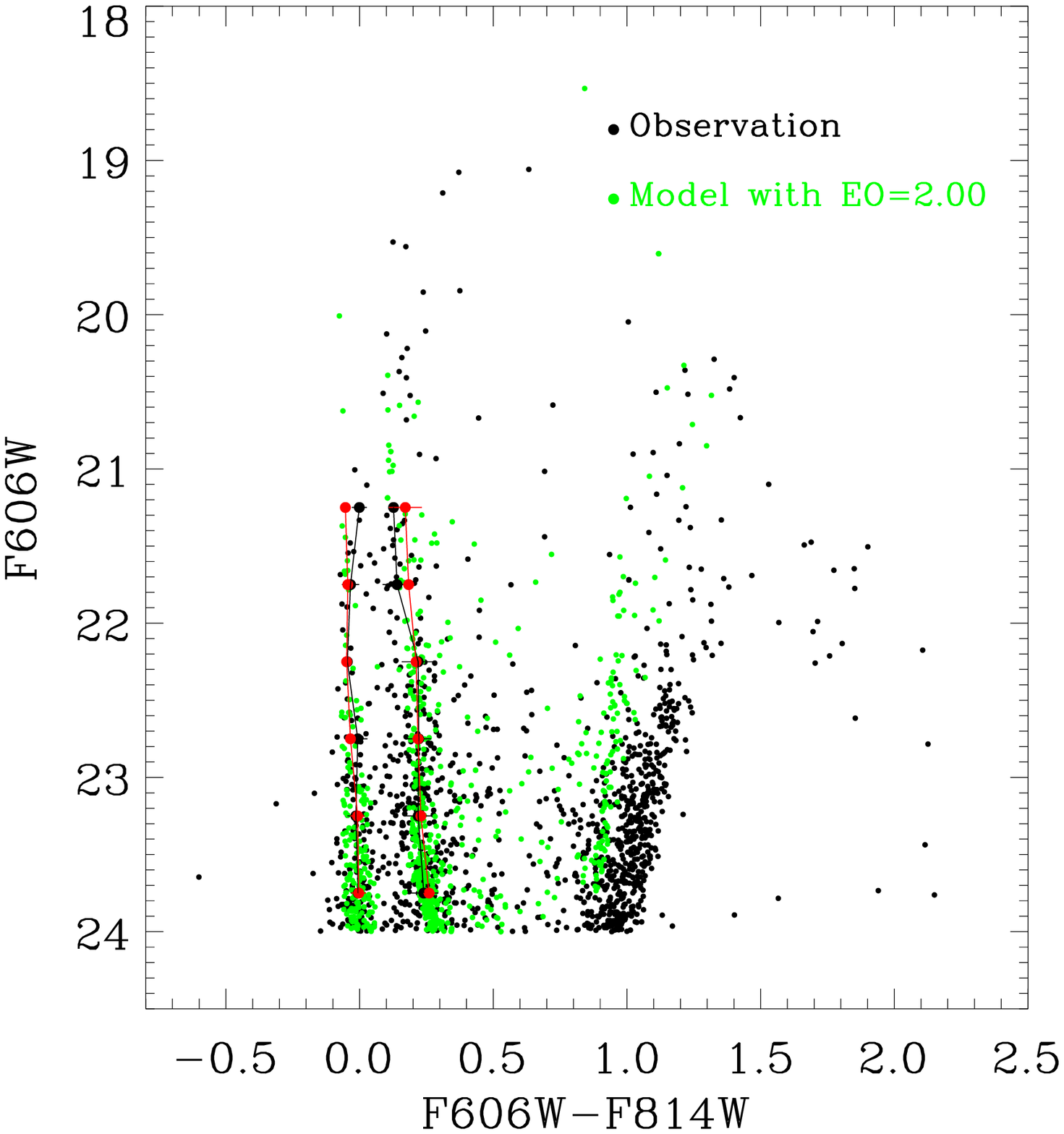}
\includegraphics[angle=0,width=0.4\textwidth]{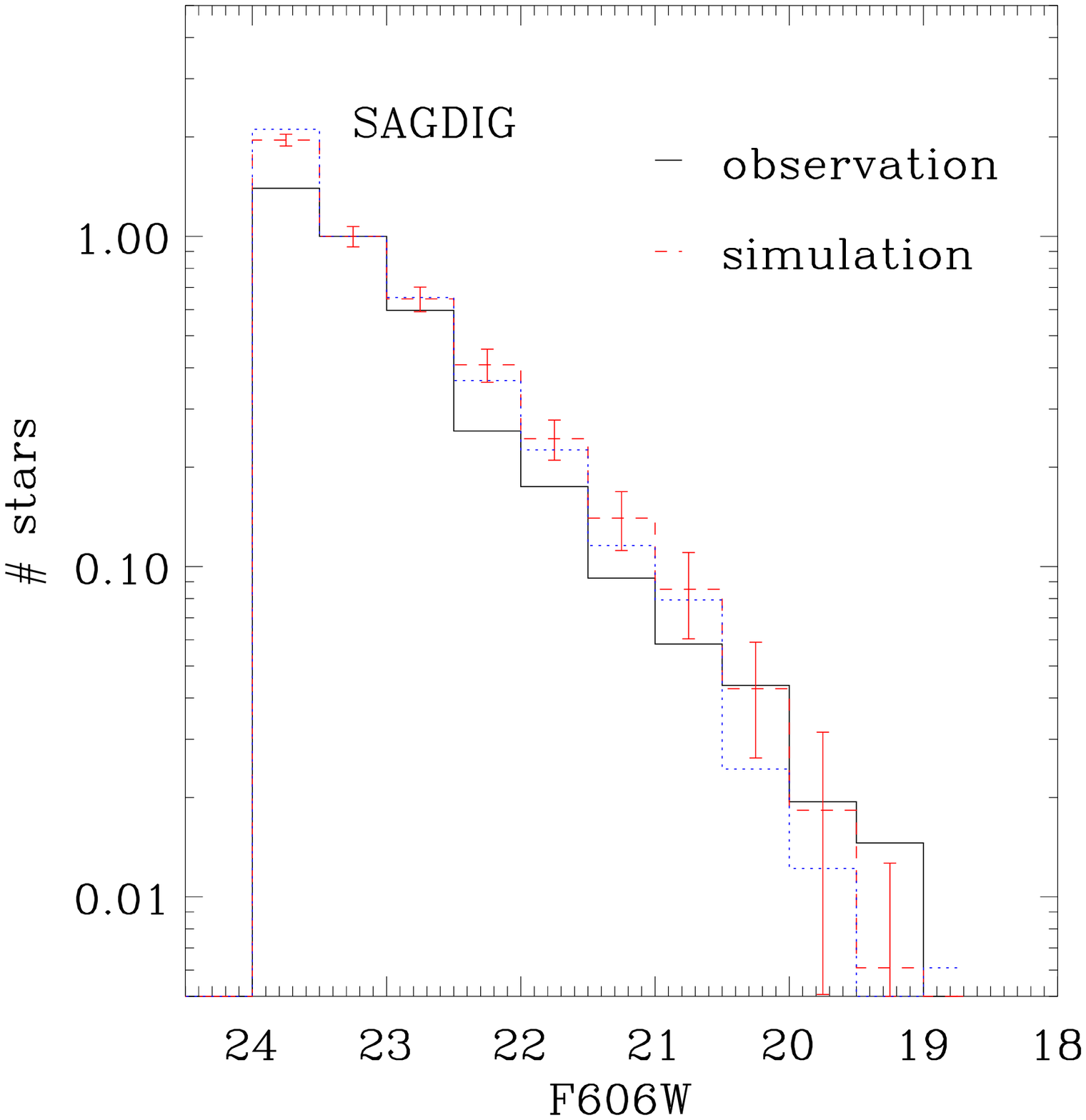}
\includegraphics[angle=0,width=0.4\textwidth]{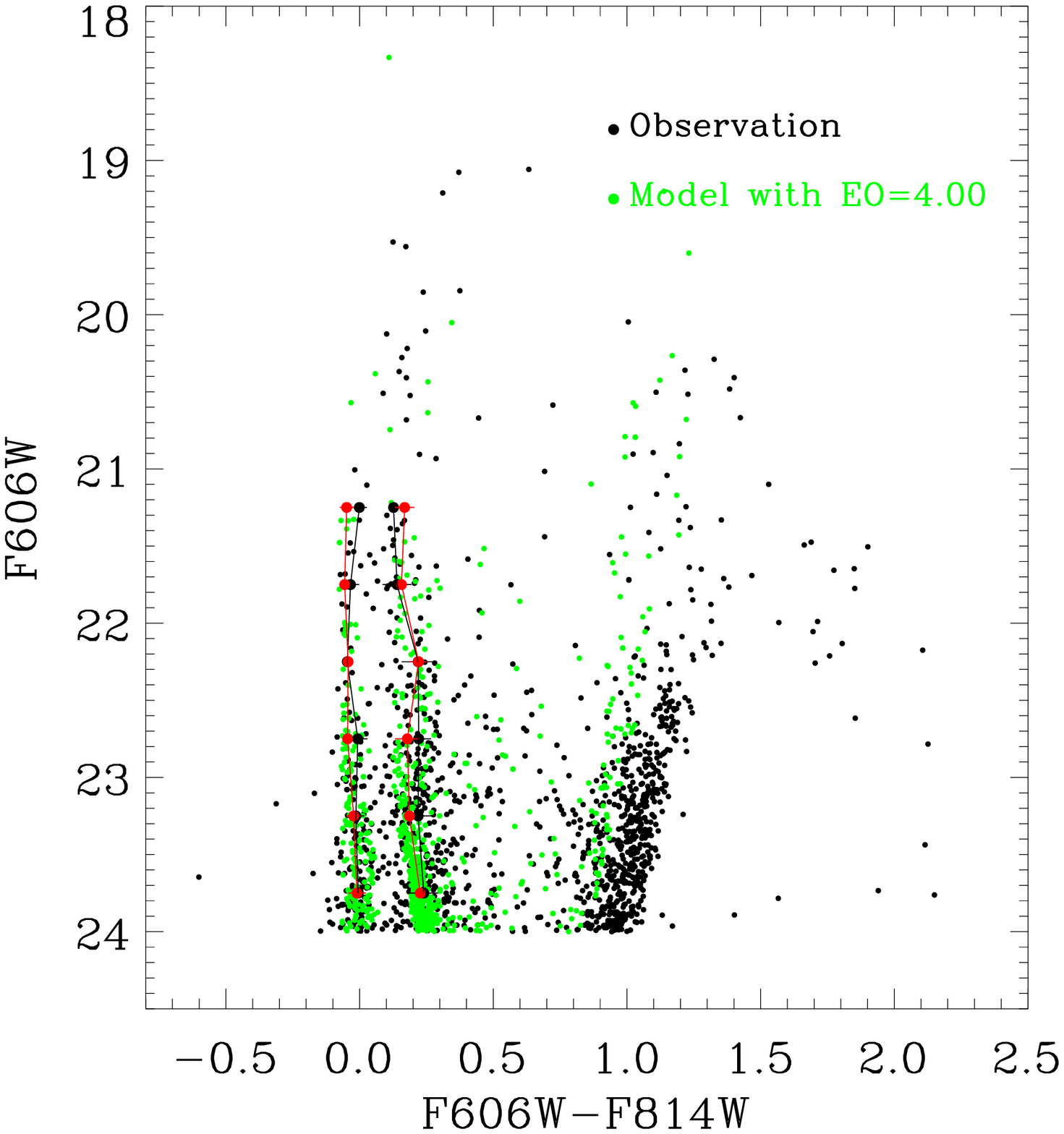}
\includegraphics[angle=0,width=0.4\textwidth]{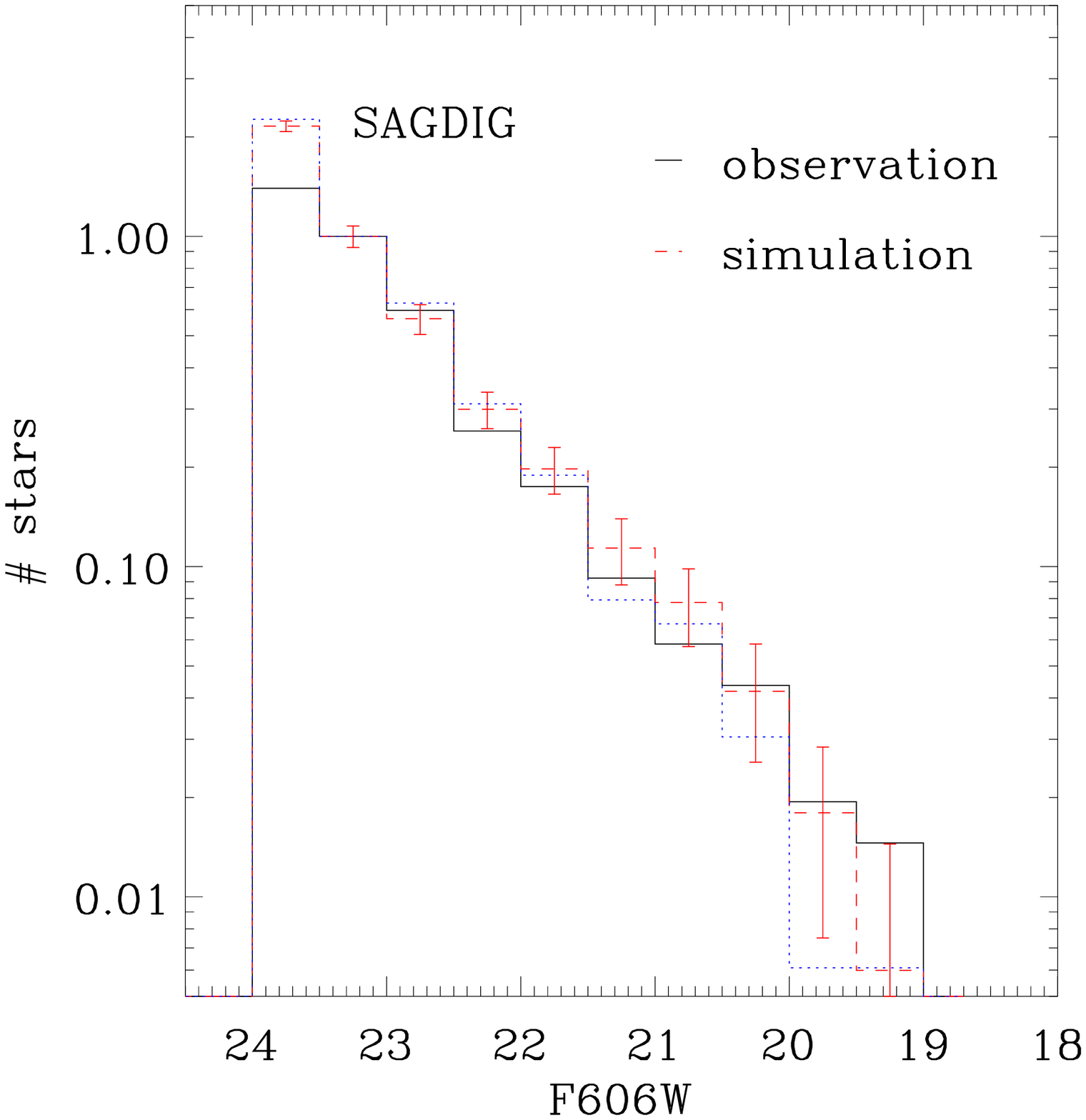}
\caption{Left panels: comparisons of observed and modelled CMDs, EO=0.7\HP, EO=2\HP and EO=4\HP, from top to bottom respectively. The observation and simulation are represented by black and green points, and the corresponding medians and fiducial lines are marked in black and red, respectively. The error bars are calculated as the median absolute deviation. Right panels: the observed (black solid) and average simulated (red dashed) luminosity functions (LFs) with vertical standard deviation bars
obtained from one hundred simulations with fixed parameters. Blue dotted lines indicate the LFs of the best-fit models. \label{fig_sagdig_sim}}
\end{figure*}
\begin{figure}
\includegraphics[angle=0,width=0.4\textwidth]{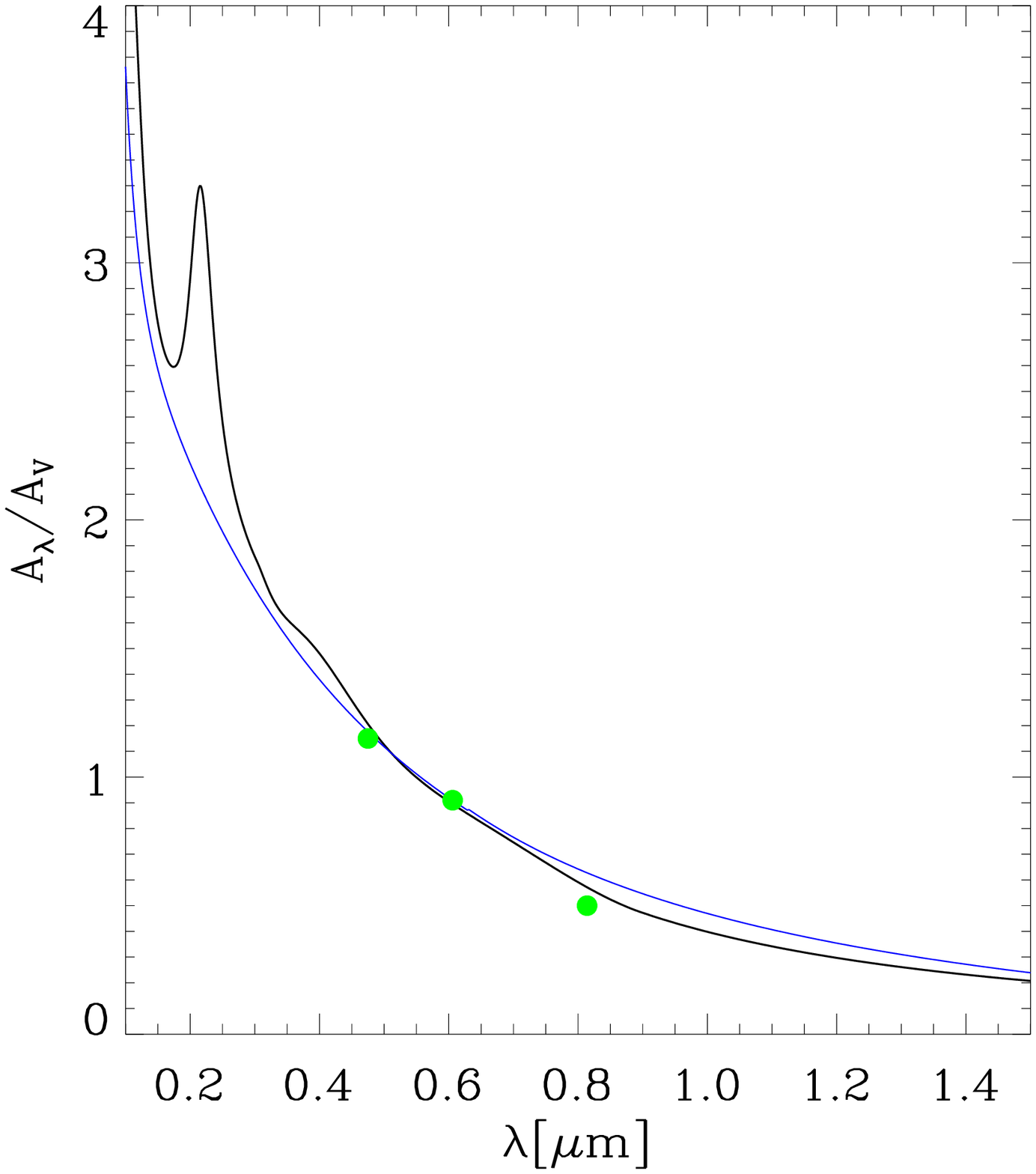}
\caption{The reddening curve A$_\lambda$/A$_V$ is shown as a function of $\lambda$. Our values A$_\lambda$/A$_V$ for different photometric bands are shown by green solid dots.
The black and blue line represent the Galactic extinction law\citep{Cardelli_etal89}
and the Calzetti Law \citep{Calzetti_etal94}, respectively. \label{ext_law}}
\end{figure}

The synthetic $m_{F606W}$ vs ($m_{F606W}-m_{F814W}$) CMDs of SagDIG obtained with the three different values of envelope overshooting are shown in Figure \ref{fig_sagdig_sim}.
The observed CMD is represented by black points, while the best simulation is shown in green color. The simulation refers to the stars brighter than $m_{F606W}$=24. For ease of comparison, we draw the observed fiducial main sequence,
which is indicated by the almost vertical black line at $(m_{F606W}-m_{F814W}) \sim 0$. It has been
obtained by considering the median of the colors of main sequence stars,
defined as the stars bluer than ($m_{F606W}$-$m_{F814W}$)=($m_{F606W}$-18)/65, in magnitude bins of $\Delta{m_{F606W}}$=0.5 in the range $21 \leq m_{F606W} \leq 24$.
The horizontal bar represents the standard deviation for the corresponding magnitude bin.
It is calculated as the median absolute deviation, which is defined as the median of an array of differences between the colors of stars and the median color. The fiducial main sequence locus  derived from the best fit CMD is shown in red color.
Since the evolution up to central H exhaustion is not affected by envelope overshooting, the main sequence loci
are the same in the three simulated CMDs.
%	a475_av=1.15; 	a606_av=0.91 ;	a814_av=0.50
%	A606=0.52
%	AV  = A606/a606_av; 	A475=a475_av * AV; 	A814=a814_av*AV
%	sag_dig_mm=25.06
The superposition of the observed and simulated main sequence loci has been obtained by assuming a single value of the extinction
for all stars in each of the three different photometric bands, A(F475W)=0.657~mag, A(F606W)=0.520~mag and A(F814W)=0.286~mag respectively. The adopted distance modulus is (m-M)$_0$=25.06.
These values of extinction and distance modulus have also been used to draw the evolutionary tracks on the CMDs in Figures \ref{fig_sagdig} and \ref{fig_sagdig_track_eo4}.
Adopting A$_{F606W}$/A$_V$=0.91 (as in the Galactic extinction law with $R_{v}$=3.1), we get A$_V$=0.57.
We compare the quantities A$_{F475W}$/A$_V$ and A$_{F814W}$/A$_V$ with
those corresponding to other typical extinction laws in Figure \ref{ext_law}.
%In the figure, our values A$_\lambda$/A$_V$ for the different photometric bands are shown by green solid dots, while the black and blue line represent the Galactic \citep{Cardelli_etal89} and the Calzetti \citep{Calzetti_etal94} extinction law, respectively.
We see that the value of A$_{F475W}$/A$_V$ lies on the Calzetti extinction curve \citep{Calzetti_etal94}, while the one of A$_{F814W}$/A$_V$ falls slightly below the Galactic one \citep{Cardelli_etal89}.
The models shown in Figure \ref{fig_sagdig_sim} correspond to the best fit selected from one hundred stochastic realizations made with the same parameters, based on the merit function that measures the agreement between the observed and modelled luminosity functions of stars both in the main sequence and in the blue-loops evolutionary phase. We notice that very bright stars ($m_{F606W} \leq 20$) in models are always much fewer than the observations. The parameters of the star formation rate and initial mass function adopted in the models are listed in Table \ref{tab_SIM_SagDIG}.
%{\color{red} errors? you adopted a single value, but there is much spread in the real population. I'd avoid claims, unless they are really significant.}

The synthetic CMDs reproduce fairly well the main features of the observed one, except for the RGB. This is because we focus on newly-formed stars, and exclude stars redder than the line ($m_{F606W}$-$m_{F814W}$)=(29.95-$m_{F606W}$)/7 which, as shown in
Figures \ref{fig_sagdig} and \ref{fig_sagdig_track_eo4}, correspond to the RGB of the old populations. We note however that the simulated green-dot sequence in the case with EO=0.7, representing
red giant stars of the intermediate age populations, seems to be more populous than
the corresponding data, while the discrepancy gets smaller
at increasing value of EO.
%{\color{red} This explains the lack of green dots in the redder part, but there is also a green-dot sequence not matched by the data (less red than the data).} {\color{blue} I think she refers to the red He-burning sequence. This sequence is close to the RGB stars, and it seems not obvious. Several factors will affect it, like differential extinction and binarity, as discussed below, and also it is relevant to the line we set, but we've already talked about it.}

As discussed in \citet{Tang_etal14}, models with enhanced envelope overshooting
produce more extended blue loops, which is
also shown in Figure \ref{fig_sagdig_track_eo4}. But the simulations show two other interesting properties. One is that, at increasing envelope overshooting, the relative fraction of blue He-burning stars increases while that of red He-burning stars decreases, explaining the better agreement
between the simulated and observed red giants obtained with EO=4.
The other is that the population of yellow He-burning stars, say those with $0.4 \leq (m_{F606W}-m_{F814W}) \leq 0.7$, decreases at increasing envelope overshooting. This is particularly evident in the magnitude range $23 \leq m_{F606W} \leq 24$.

The effect of envelope overshooting on the extension of the blue loops can be appreciated already by eye from Figure \ref{fig_sagdig_sim}, but in order to render it more clear,
we derive the BHeB stars main locus by using the same method adopted for the main sequence locus.
The bins are the same used for the main sequence stars, though we recall that, at fixed initial mass,
He-burning stars are brighter than H-burning ones. In order to derive the median color we have considered
all stars redder than ($m_{F606W}$-$m_{F814W}$)=($m_{F606W}$-18)/65 and bluer than ($m_{F606W}$-$m_{F814W}$)=0.4.
The meaning of the error bar is the same as the one obtained for MS stars.
The median locus of the observed BHeB stars is drawn in black and it runs almost parallel to the
main sequence locus, but about 0.25~mag redder. The locus of the synthetic BHeB stars is drawn in red.
For EO=0.7\HP, it is 0.1~mag redder than the observed one at magnitudes $23 \leq m_{F606W} \leq 24$. At brighter magnitudes
the difference disappears. The locus of the models with envelope overshoot EO=2\HP
runs superimposed to that of the observed data, in the magnitude range $23 \leq m_{F606W} \leq 24$, while
using the models with envelope overshoot EO=4\HP, it is slightly bluer than the observed one,
in the same magnitude range $23 \leq m_{F606W} \leq 24$.
We note that the difference between the observed and modelled BHeB loci is not large, even in the case of EO=0.7\HP. Actually
in the latter case the largest difference is comparable to the standard deviations of the loci themselves. However if we base
our judgment  more on the systematics of the effect than on its entity, it is clear that only models with larger envelope overshooting are able to reproduce the extension of the blue loops in the magnitude range $23 \leq m_{F606W} \leq 24$.

\subsection{The star formation rate}
As in \citet{Tang_etal14}, the star formation rate is represented by an exponential parametrization
\begin{equation}
   SFR(t) = SFR_0\times exp(\frac{t}{\tau})
\end{equation}
where $t$ represents the stellar age, SFR$_0$ the current value of the SFR and $\tau$ the characteristic $e$-folding time.
The results are shown in Table \ref{tab_SIM_SagDIG}. We find that the SFR in \object{SagDIG} increases toward recent times ($\tau < 0$). Considering the case of EO=2\HP, the average SFR in the last 100~Myr ($\sim$9E-4~M$_\odot$/yr) is significantly lower than that we obtained for  \object{Sextans A} ($<$SFR$>$=2.9E-3~M$_\odot$/yr), \object{WLM} (($<$SFR$>$=2.7E-3~M$_\odot$/yr)) and \object{NGC 6822} (($<$SFR$>$=3.7E-3~M$_\odot$/yr)). It is worth noting that this average SFR is for the whole galaxy, while in those three galaxies the derived SFR refer to selected star-forming regions.
We also notice that the SFR does not change much for models with different envelope overshooting, but in the case with high envelope overshooting, EO=4\HP,
the SFR turns out to be $\sim$ 30\% larger than that in the case of EO=0.7\HP. Correspondingly, the mass formed in the recent burst
amount to  M*=9.26E5\Msun with EO=4\HP, while it is relatively smaller in the other two cases, M*$\sim$7E5\Msun.
Our value is close to that of \citet{Karachentsev_etal99} who found
$<$SFR$>$=6.6$\pm$0.8E-4\Msun/yr in the age range 0.05-0.2~Gyr. Indeed, in the same period, we find $<$SFR$>$=8.2E-4\Msun/yr using models with EO=0.7\HP.
In Table \ref{tab_SIM_SagDIG} we also show  the masses of the most massive and  of the brightest stars found in the simulations.
%1.26E6 9E5

\subsection{Effects of differential extinction and binary stars}
A remarkable property of the  CMDs shown in Figure \ref{fig_sagdig_sim}
is that the standard deviation of the observed fiducial sequences are larger than
the modelled ones. This indicates that both the observed main sequence and the
observed BHeB sequence are more dispersed than that predicted by the models,
at least for stars brighter than $m_{F606W}$=24. This cannot be ascribed
to photometric errors because, besides being explicitly included in the simulation,
they are by far too small to explain the effect. Thus other explanations have to be found.

In the case of the main sequence, the discrepancy could be due to the well known {\sl main sequence widening} effect, i.e. that the termination point of the observed main sequence is cooler than
that predicted by the models. This could be appreciated in the comparison between the observed
CMD and the evolutionary tracks in Figure \ref{fig_sagdig_track_eo4} where the width of the main sequence
is marked by the pink lines. This discrepancy is one of the motivations that
inspire the presence of extended mixing effects during the central H-burning phase
of intermediate- and high-mass stars \citep{Massevitch_etal79,Bressan_etal81}.
In the case of the BHeB stars, the problem does not have a similar explanation.
The star number counts in the evolved phases are proportional to the evolutionary lifetimes
of the corresponding phases, and to explain a large dispersion one should invoke a mechanism that
is able to slow down the transition from the RSG to the BSG phases which is at present not known.

There are two other effects that could explain the widening of the sequences.

\begin{figure}
\includegraphics[angle=0,width=0.4\textwidth]{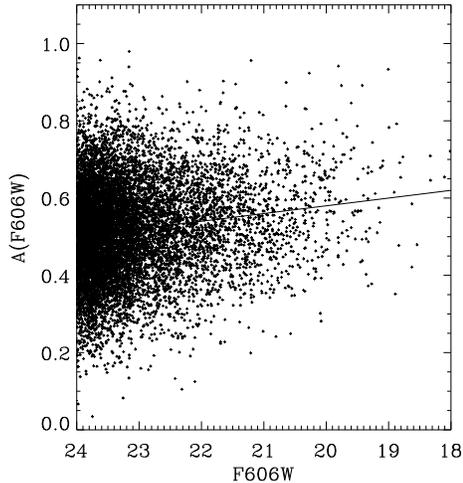}
\caption{Differential extinction is assumed as a certain dispersion around a mean value A(F606W)=-0.02 $\times$ $m_{F606W}$+0.98 (black line). \label{fig_attenuation}}
\end{figure}

\begin{figure}
\includegraphics[angle=0,width=0.4\textwidth]{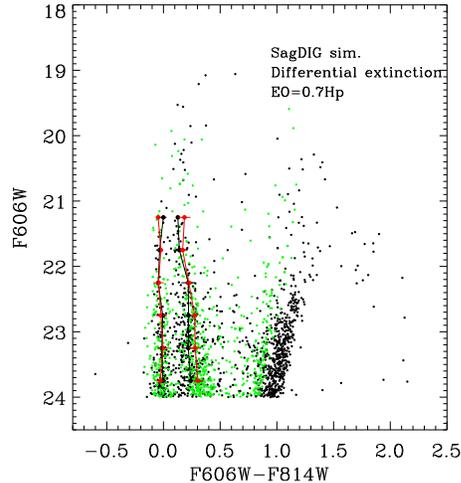}
\includegraphics[angle=0,width=0.4\textwidth]{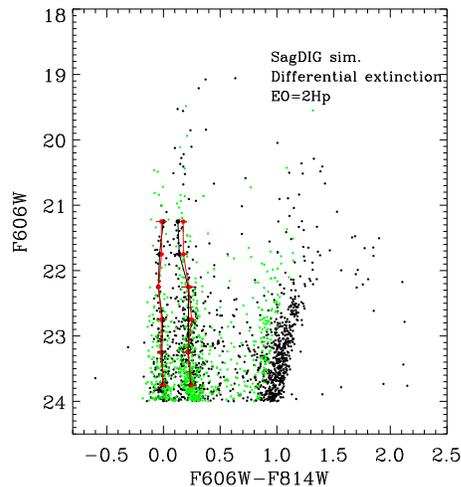}
\includegraphics[angle=0,width=0.4\textwidth]{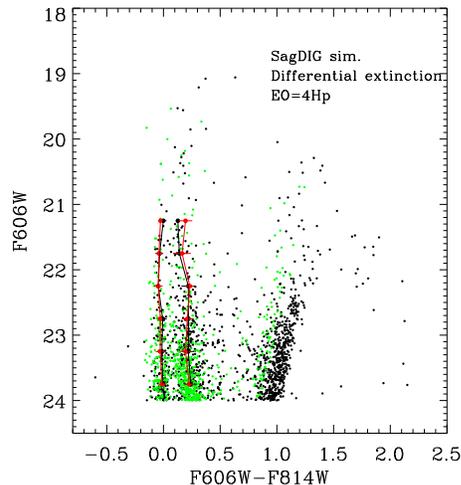}
\caption{The best model obtained by {\sl\,PARSEC}~V1.2 with EO=0.7\HP, 2\HP and 4\HP from top to bottom with differential extinction. \label{fig_sagdig_ext}}
\end{figure}

One is differential extinction. We have already mentioned that in previous analysis of DIGs, \citet{Bianchi_etal12} and \citet{Tang_etal14} have directly measured and then modelled differential extinction of individual stars.
This effect certainly contributes to the widening of both the main sequence and the blue He-burning sequence and thus helps filling the gap between them.
But unfortunately, at variance with the quoted DIGs, in the case of SagDIG we lack the broad multiband
photometry which allows \citet{Bianchi_etal12} to obtain estimates of attenuation for individual stars.
Nevertheless we may try to estimate the size of this effect, by assuming that the models
are correct. Applying the same method used in \citet{Tang_etal14}, we derive a more realistic estimate of the extinction of individual stars,
according to the trend of increasing attenuation at increasing luminosity, as shown in the diagram
A(F606W) vs F606W of Figure \ref{fig_attenuation}.
Adopting the differential extinction we obtain the simulated CMDs shown in Figure \ref{fig_sagdig_ext}.
While the gap can be partly filled by introducing the differential reddening, it remains evident in the case of EO=0.7\HP.
The gap decreases in the case of EO=2\HP and almost disappears when a high envelope overshooting is adopted, EO=4\HP.
The fiducial lines of the main sequence and of the BHeB stars, drawn in the same way as in Figure \ref{fig_sagdig_sim},
show that the fact that the models with the standard value of envelope overshooting, EO=0.7\HP, are not able to reproduce the observed extended loops during central He-burning phase, is not due to neglecting differential extinction.
Even with differential extinction a large value of envelope overshooting (EO$\sim$2.0) is favoured.
\begin{table}
\begin{center}
\caption{Parameters of the CMD simulations of SagDIG \label{tab_SIM_SagDIG}}
\begin{tabular}{ccccccc}
\hline\hline
EO & $\tau$ & $\alpha$ & SFR$_0$ & $<$SFR$>$\textsuperscript{*} & M$_{max}$& M$_{bright}$\\
 & {\rm yr} & & {\rm M$_\odot$/yr} & {\rm M$_\odot$/yr}& {\rm M$_\odot$}& {\rm M$_\odot$}\\
\hline
0.7 &  -2E9 & 2.05 &  8.79E-4&  8.57E-4 & 39 & 11\\
2   &  -2E9 & 2.25 &  9.65E-4&  9.41E-4 & 62 & 13\\
4   &  -2E9 & 2.15 &  1.18E-3&  1.15E-3 & 30 & 30\\
\hline
\multicolumn{4}{l}{\textsuperscript{*}\footnotesize{the average SFR in the last 100~Myr}}
\end{tabular}
\end{center}
\end{table}

Another effect that may contribute to widen the theoretical nominal sequences of single stars is
the presence of binary stars.
%As is well known, the incidence of binaries is high, of the order of 50\% \citep{Duquennoy_etal91}.
%The binary fraction appears higher for more massive stars, reaching 70\% \citep{Abt_etal90}
% and decreasing  to 40\% for M dwarfs \citep{Kroupa_etal93}.
As is well known, the incidence of binarity in stars is high. \citet{Raghavan_etal10} analysed the sample of stars from the {\it Hipparcos} catalogue, suggesting 33\% of solar-type stars in the solar neighbourhood are binaries. Moreover, young massive stars are believed more likely to be binaries. \citet{Sana_etal09} studied the optical spectra of O-type stars in NGC 6611 and derived the minimal binary fraction to be 0.44, but it could be increased up to 0.67 if all binary candidates are confirmed.

\begin{figure}
\includegraphics[angle=0,width=0.4\textwidth]{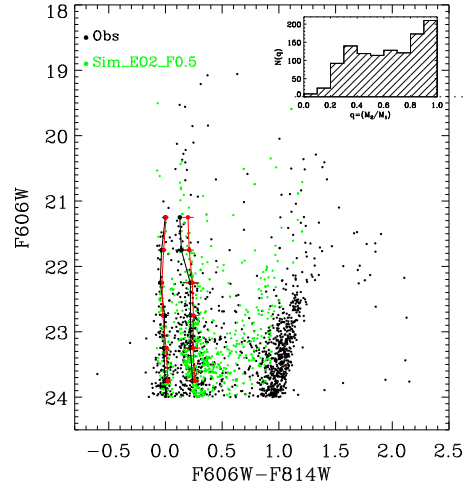}
\caption{The best model obtained by {\sl\,PARSEC}~V1.2 with EO=2\HP and binary fraction F=0.5. The insert shows the mass ratio distribution of binaries.  \label{sim_binary}}
\end{figure}

In order to estimate the effect of binaries on synthetic CMDs, we run a model with a percentage of binaries of 50\%.
%{\color{red} what mass ratio range?} {\color{blue} We don't assume the distribution in advance, and later we show the distribution. no correction here.}
%{\bf [Note that the binaries we discuss here are not only composed by really physical bound systems but also include possible multiple stars aligned along the same line of sight.]}
%{\color{red} Unless necessary, this could be omitted.}{\color{blue} So delete?}
To reproduce the assumed 50\% contamination, we randomly combine the sample stars in the mock catalogue without assuming a particular value of the mass ratio,
until we reach the total number of observed objects with the required binary fraction.
Since for binaries consisting of equal-mass components the apparent magnitude may increase by up to 0.75~mag, to obtain a complete estimate of their effect above $m_{F606W}$=24, we
first work on stars brighter than $m_{F606W}$=25. After taking into account the luminosity increase due to binarity, we further select objects brighter than $m_{F606W}$=24 and bluer than the line ($m_{F606W}$-$m_{F814W}$)=(29.95-$m_{F606W}$)/7, for stars redder than this line are considered as RGB stars.
This procedure is repeated one hundred times to obtain the best-fit model, the average luminosity function and its standard deviation at the selected magnitude bins.
In this case we adopt a fixed extinction, as discussed in the previous section.
Figure \ref{sim_binary} shows the simulated CMD obtained by the models with EO=2\HP and a binary fraction of 50\%. The inset shows the distribution of the mass ratios $q=M_{2}/M_{1}$ between the primary and secondary components,  which is consistent with a flat distribution \citep{Mermilliod_etal92}.

Compared to the middle panel of Figure \ref{fig_sagdig_sim}, we see the effect of binaries is to broaden the main sequence and also the BHeB sequence. The number of stars falling between these two sequences
is larger than that in the case computed without considering binaries.
However the BHeB sequence is clearly split into two parallel sequences
between $23 \leq m_{F606W} \leq 24$, and the red He-burning sequence seems also too broad, as part of these stars move to the location of yellow giant/supergiant. In particular the splitting of the BHeB seqence is not seen in the observed diagram, perhaps indicating that either the assumed binary fraction or the resulting mass ratio are too high for this galaxy of low metallicity.
Furthermore, because of the asymmetric behaviour of the superposition of star pairs in the CMD,
the fiducial lines shift toward the red side and,
in order to reconcile the model with the observations, one should make use of a lower
attenuation, by a factor $\delta(m_{F606W}-m_{F814W})\sim$0.02.
%{\color{red} strange! so ... no binaries? or no flat mass ratio?}

\section{Discussion and Conclusion}\label{sec:discon}
In \citet{Tang_etal14} we had obtained strong indications that,
in presence of sizable overshooting from the convective core during central H burning,
the standard {\sl\,PARSEC} models with envelope overshooting EO=0.7\HP fail to reproduce the width of the observed blue loops of intermediate- and high-mass stars
in three star forming dwarf galaxies in the Local Group, \object{Sextans A}, \object{WLM} and \object{NGC6822}, characterized by low metallicity, 0.001$\leq$Z$\leq$0.004.
Meanwhile we found that a significant envelope overshooting at the bottom of the convective envelope, EO=2.0-4.0\HP,
must be used to overcome this discrepancy in models with metallicity typical of the aforementioned galaxies, 0.001$\leq$Z$\leq$0.004.
We stressed that this discrepancy could be cured by adopting a metallicity significantly lower than the observed one,
which however indicates that using the extension of the blue loops predicted by  models to infer the metallicity
of the galaxies could be risky.
In this paper we continue this investigation by considering the case of SagDIG, a nearby star forming dwarf irregular whose metallicity is estimated to be
even lower than that of the aforementioned galaxies, Z=0.0005.
This galaxy is an ideal workbench to test the performance of models of intermediate- and high-mass stars because it harbours a recent burst of star formation, and
it is sufficiently nearby that intermediate-mass stars with masses as low as M=2\Msun can be detected. The foreground contamination has been eliminated using proper motions of individual stars \citep{Momany_etal14}.

From a preliminary superposition of the standard {\sl\,PARSEC} evolutionary tracks with EO=0.7\HP and the observed CMD of SagDIG, we already see that the models are not able to reproduce the observed loops. This has already been noticed by \citet{Momany_etal05} who, in an attempt to determine the metallicity of the galaxy from the blue loop superposition, were forced to try also the lowest
value of metallicity of the old Padova models, Z=0.0001.

Before performing new calculations with enhanced envelope overshooting, as suggested by \citet{Tang_etal14},
we discuss if the problem could be alleviated by adopting the Ledoux criterion instead of the Schwarzschild
one for the determination of the unstable region. On one hand, it has already been shown in the past that the adoption of the Ledoux criterion may favour
the development of extended blue loops \citep{Chiosi_Summa70}.
On the other hand, the fact that the problem of short blue loops is found even in the tracks of intermediate-mass stars, for which
the results should be independent of the adopted instability criterion because they
do not possess intermediate unstable region (within the profile of  chemical composition),
suggests that the reason should be different.
By analyzing a model of M=12\Msun and Z=0.004 under standard {\sl\,PARSEC} assumptions, which evolves without performing a blue loop, we find that the Ledoux criterion tend to {\it suppress} the blue loop,
in the sense that the star favours the red-ward evolution after central H burning.
This is apparently in contrast to the results reported by \citet{Chiosi_Summa70}, who indicated that the Ledoux criterion favours the blue loop.
However, a thorough inspection of that seminal paper shows that their model with the Schwarzschild criterion does not perform a blue loop only because
it begins central He burning already in the blue loop region, i.e. as a BHeB star.
With a series of new models not reported here for seek of conciseness,
we show that at He ignition, the mass pocket between the H-exhausted core and the H/He discontinuity is
already negligible, fully confirming the finding of \citet{Walmswell_etal15}.
The Schwarzschild criterion allows the formation of a larger intermediate convective shell than
in the case of the Ledoux criterion, and this convective shell pushes the H/He discontinuity deeper
in the star, reducing the above mass pocket.
Thus the case A model in \citet{Chiosi_Summa70} begins He burning already in the BHeB equilibrium configuration and not in the RSG stage, without the need of performing a blue loop.
On the contrary, their case B model, with the Ledoux criterion, encounters the condition of a thin mass pocket between the H-exhausted core and the H/He discontinuity,
slightly later during the evolution, after He ignition in the RSG phase, and thus performs a blue loop.
When significant core overshooting is allowed during the H-burning phase, the situation changes critically.
Being the H-exhausted core larger than that in the case without core overshooting,
its contraction after central H-burning is stronger,
pushing the star into the RSG stage, independently of the adopted instability criterion.
These are the two cases shown in Figure \ref{convection_criteria} for M=12\Msun and Z=0.004.
However,  though the differences between the models computed with the two different criteria are minimal,
the track computed with the Schwarzschild criterion develops a larger intermediate
convective region that deepens slightly the H-He discontinuity than in the one computed with the
Ledoux criterion. The former develops the blue loop while in the latter case
the condition of a thin intermediate layer is encountered only
toward the end of central He burning, when the BHeB structure is no more
a possible configuration for the star so that
it burns the entire central He in the RSG stage.

Having definitely excluded the Ledoux criterion as a possible cure of the problem of the loops extension,
we computed additional evolutionary tracks with larger values of  envelope overshooting,
EO=2\HP and  EO=4\HP. Combining the results with other specified parameters, the IMF and SFR law, the photometric errors and the extinction in our own CMD simulator,
we construct the synthetic CMDs that are compared with the observed one of SagDIG.
In all models the location of the observed fiducial main sequence is well reproduced with a reasonable value of the
attenuation. This is expected since the main sequence phase is not affected by envelope overshooting
and the match is actually used to determine the extinction to be adopted in the simulated CMDs.
As expected, models with larger envelope overshooting perform more extended blue loops, and their BHeBS get closer to the MS.
In order to decide which value of envelope overshooting reproduces better
the observed extension of the blue loops, we also draw and compare the fiducial lines corresponding to the BHeBS of the synthetic  and of the observed CMDs.
We find that the model with EO=2\HP matches the observations best, while the blue loops predicted by the case EO=0.7\HP are not hot enough and those predicted
in the case of EO=4\HP are likely hotter than the data indicate.
It is worth stressing that, since the value of the envelope overshooting is not an adjustable parameter in the CMD comparison,
we have no better and more statistically sound method to decide which is the best case, than that of comparing the fiducial main sequences.
We have also tested how the results depend on other additional assumptions concerning the attenuation and the possible effect of binarity.
If we assume a differential attenuation with a  reasonable model (Figure \ref{fig_attenuation}) where the attenuation has a certain dispersion around a mean value
that increases with the intrinsic luminosity of the stars \citep{Tang_etal14}, the overall CMD fit looks better but the
preferred value of the envelope overshooting remains unchanged.
However, due to the dispersion introduced in the synthetic CMD, the uncertainty of the fiducial points (the horizontal bars in the CMD figures), corresponding to
the color median absolute deviation in each magnitude bin, becomes slightly larger.
To single out the effect of binarity, we consider
the case with constant attenuation and
 a binary fraction of 50\%.
The effect of binaries is to broaden both the main sequence and the BHeB sequence, with the consequence that
the number of stars falling between the two sequences
is larger than that in the case computed without considering binaries.
However the BHeB sequence is clearly split into two parallel sequences
between $23 \leq m_{F606W} \leq 24$, and the red He-burning sequence becomes less populated, as part of these stars move into the region populated by yellow giants/supergiants.
The splitting of the BHeB sequence is not seen in the observed diagram, perhaps indicating that either the assumed binary fraction or the resulting mass ratio are too high for this galaxy of low metallicity.
Furthermore, because of the asymmetric behaviour of the superposition of star pairs in the CMD,
the fiducial lines shift toward the red side and, in order to reconcile the model with the observations, one should adopt a slightly lower
attenuation, $\delta(m_{F606W}-m_{F814W})\sim$0.02.
Even in this case the models that perform better are those computed with an envelope overshooting EO=2\HP.

The results of this paper are consistent with those found by \citet{Tang_etal14} for three galaxies with slightly higher
metallicities than the one considered here.
Thus the current investigation corroborates the finding that the mixing scale below the formal Schwarzschild border
in the envelopes of intermediate- and high-mass stars must be significantly higher than currently assumed in {\sl\,PARSEC} models.
We recall that there are other conditions where
strong mixing below the conventional instability region is invoked
such as, for example, to enhance the efficiency of the carbon dredge up during the thermally pulsing Asymptotic Giant Branch phase \citep{Kamath_etal12}.
On the other hand, we note that such a high value is likely incompatible with
the location of the Red Giant Branch (RGB) bumps observed in Globular Clusters,
which can be well reproduced by the models using an envelope overshooting not larger than EO=0.5\HP.
Thus, if the interpretation in terms of mixing into the stable regions due to convection is correct, it is likely  that the mixing scale below the formal Schwarzschild border is not a fixed
fraction of the pressure scale height, as assumed in usual extensions of the MLT, but instead it depends on the interior structure of the star.
In fact our results highlight a difficulty inherent in the MLT
which, by itself, cannot predict the size of the velocity field.
%On the other hand in spite of the huge efforts made with hydrodynamical simulations
%there is not yet  a theory suited for massive evolutionary track computations (e.g. Arnett et al 2015).
In this respect it is interesting to compare our finding with the results of a recent 3D implicit large eddy simulation of the turbulent convection in the envelope of a 5\Msun red giant star \citep{Viallet_etal13}.
The simulation refers to the envelope of a 5\Msun star at the end of central He burning.
The convective unstable region extends inward from an outer radius R$\sim$4.1$\times$10$^{12}$cm to an inner radius r$\sim$2.2$\times$10$^{12}$cm, where the Schwarzschild 
condition of stability is satisfied ($\nabla_{rad}=\nabla_{ad}$, see their Figure 17). 
At this layer, the velocity field shows a sudden drop, but does not vanishes; it ceases
to be dominated by the radial components and begins to be dominated by the transversal components (see their Figure 4). This indicates a clear regime of convective turn-over extending 
for about 0.6\HP ($\sim$0.25$\times$10$^{12}$cm)
below the formally unstable region, which the authors recognize as the lower {\it boundary} of the convective envelope. In this region the rms velocities decrease from $\sim$10$^5$cm/s to $\sim$10$^4$cm/s. 
Thus the simulation shows that there is a region of about 0.6\HP below the formal Schwarzschild border
where velocities remain significantly high, characterized by the transversal components being about twice the radial ones. From Figure 4 of \citet{Viallet_etal13} one may also see that the velocity field extends much below this point 
and, though it is difficult to evaluate from the figure alone,
the rms velocities are clearly larger than $\sim$10$^3$cm/s, even at several $\sim$10$^{11}$cm
below the Schwarzschild border.

In our track of M=5\Msun and Z=0.004, at the stage of maximum  penetration
near the tip of the first red giant ascent (not the second as in the \citet{Viallet_etal13} simulation), the model has the following structure. The stellar radius is R$\sim$7.4$\times$10$^{12}$cm 
and the inner Schwarzschild border is at r$\sim$0.8$\times$10$^{12}$cm.  
The size of the unstable region is $\sim$6.6$\times$10$^{12}$cm.
The maximum velocities predicted using the MLT are a few $\sim$10$^5$cm/s,
but drop to $\sim$2.4$\times$10$^4$cm/s near the Schwarzschild border.
The pressure scale height at the inner Schwarzschild border is \HP$\sim$4$\times$10$^{11}$cm
and a mixing scale of 0.7\HP corresponds to $\delta_r(0.7)\sim$2.4$\times$10$^{11}$cm while,
one of 4\HP corresponds to $\delta_r(4)\sim$6$\times$10$^{11}$cm.
This non linearity is due to the fact that we measure the overshooting region in terms of
pressure and not in length, and that the pressure scale height below the bottom of the convective region decreases at decreasing radius (while it keeps increasing again in the inner He core).
Our convective region is thus not very different from the one described in \citet{Viallet_etal13}
simulation. With radial velocities of about $\sim$10$^3$cm/s, convection can span
a distance of  $\delta_r(4)$ in about 20yrs which,  
since our model spends a few thousand years before convection begins to retreat,
is about 100 times shorter than the evolutionary time of the star in this phase.
Thus, at first glance, the \citet{Viallet_etal13} simulation indicates that the internal mixing
of about 2-4\HP below the formal convective border are possible
in an evolutionary timescale.
What we have neglected in this simple discussion are the effects of the
molecular weight discontinuity that forms between the outer H rich and  
the inner He rich regions, and that could strongly limit or even suppress the penetration of convective motions (e.g. \citet{Bressan_etal81}). 
At face value, our finding indicates that turbulent entrainment into the stably stratified layers \citep{Meakin_arnett07} should be quite efficient in the envelopes of such stars.
Clearly detailed numerical simulations (e.g. \citet{Arnett_etal15}) especially addressed to the analysis of the mixing efficiency in presence of chemical composition gradient, would be extremely helpful to clarify this issue.

\section{Acknowledgements}
We thank S. Charlot and   Y. Chen for helpful discussions.
A.B. acknowledges support from INAF through grant PRIN-2014-14.
P.M. acknowledges support from {\em Progetto di Ateneo 2012}, University of Padova,  ID: CPDA125588/12.

%%%%%%%%%%%%%%%%%%%%%%%%%%%%%%%%%%%%%%%%%%%%%%%%%%%%%%%%%
%to be commented before sending to editor
\bibliographystyle{mn2e_new} %style mn.bst
\bibliography{sagdig} % your references file.bib

\begin{thebibliography}{39}
\expandafter\ifx\csname natexlab\endcsname\relax\def\natexlab#1{#1}\fi

\bibitem[{{Alongi} {et~al}\mbox{.}(1991){Alongi}, {Bertelli}, {Bressan}, \&
  {Chiosi}}]{Alongi_etal91}
{Alongi} M., {Bertelli} G., {Bressan} A., {Chiosi} C., 1991, \aap, 244, 95

\bibitem[{{Arnett} {et~al}\mbox{.}(2015){Arnett}, {Meakin}, {Viallet},
  {Campbell}, {Lattanzio}, \& {Moc{\'a}k}}]{Arnett_etal15}
{Arnett} W.~D., {Meakin} C., {Viallet} M., {Campbell} S.~W., {Lattanzio} J.~C.,
  {Moc{\'a}k} M., 2015, \apj, 809, 30

\bibitem[{{Asplund} {et~al}\mbox{.}(2009){Asplund}, {Grevesse}, {Sauval}, \&
  {Scott}}]{Asplund_etal09}
{Asplund} M., {Grevesse} N., {Sauval} A.~J., {Scott} P., 2009, \araa, 47, 481

\bibitem[{{Bertelli} {et~al}\mbox{.}(1990){Bertelli}, {Betto}, {Chiosi},
  {Bressan}, \& {Nasi}}]{Bertelli_etal90}
{Bertelli} G., {Betto} R., {Chiosi} C., {Bressan} A., {Nasi} E., 1990, \aaps,
  85, 845

\bibitem[{{Bertelli}, {Bressan} \& {Chiosi}(1985){Bertelli}, {Bressan}, \&
  {Chiosi}}]{Bertelli_etal85}
{Bertelli} G., {Bressan} A.~G., {Chiosi} C., 1985, \aap, 150, 33

\bibitem[{{Bianchi} {et~al}\mbox{.}(2012){Bianchi}, {Efremova}, {Hodge},
  {Massey}, \& {Olsen}}]{Bianchi_etal12}
{Bianchi} L., {Efremova} B., {Hodge} P., {Massey} P., {Olsen} K.~A.~G., 2012,
  \aj, 143, 74

\bibitem[{{B{\"o}hm-Vitense}(1958)}]{Bohm-Vitense58}
{B{\"o}hm-Vitense} E., 1958, \zap, 46, 108

\bibitem[{{Bressan} {et~al}\mbox{.}(2013){Bressan}, {Marigo}, {Girardi},
  {Nanni}, \& {Rubele}}]{Bressan_etal13}
{Bressan} A., {Marigo} P., {Girardi} L., {Nanni} A., {Rubele} S., 2013, in
  European Physical Journal Web of Conferences, Vol.~43, European Physical
  Journal Web of Conferences, p. 3001

\bibitem[{{Bressan} {et~al}\mbox{.}(2012){Bressan}, {Marigo}, {Girardi},
  {Salasnich}, {Dal Cero}, {Rubele}, \& {Nanni}}]{Bressan_etal12}
{Bressan} A., {Marigo} P., {Girardi} L., {Salasnich} B., {Dal Cero} C.,
  {Rubele} S., {Nanni} A., 2012, MNRAS, 427, 127

\bibitem[{{Bressan}, {Chiosi} \& {Bertelli}(1981){Bressan}, {Chiosi}, \&
  {Bertelli}}]{Bressan_etal81}
{Bressan} A.~G., {Chiosi} C., {Bertelli} G., 1981, \aap, 102, 25

\bibitem[{{Calzetti}, {Kinney} \& {Storchi-Bergmann}(1994){Calzetti}, {Kinney},
  \& {Storchi-Bergmann}}]{Calzetti_etal94}
{Calzetti} D., {Kinney} A.~L., {Storchi-Bergmann} T., 1994, \apj, 429, 582

\bibitem[{{Cardelli}, {Clayton} \& {Mathis}(1989){Cardelli}, {Clayton}, \&
  {Mathis}}]{Cardelli_etal89}
{Cardelli} J.~A., {Clayton} G.~C., {Mathis} J.~S., 1989, \apj, 345, 245

\bibitem[{{Chen} {et~al}\mbox{.}(2014){Chen}, {Girardi}, {Bressan}, {Marigo},
  {Barbieri}, \& {Kong}}]{Chen_etal14}
{Chen} Y., {Girardi} L., {Bressan} A., {Marigo} P., {Barbieri} M., {Kong} X.,
  2014, \mnras, 444, 2525

\bibitem[{{Chiosi} \& {Summa}(1970)}]{Chiosi_Summa70}
{Chiosi} C., {Summa} C., 1970, \apss, 8, 478

\bibitem[{{Demers} \& {Battinelli}(2002)}]{Demers_etal02}
{Demers} S., {Battinelli} P., 2002, \aj, 123, 238

\bibitem[{{Girardi}, {Rubele} \& {Kerber}(2009){Girardi}, {Rubele}, \&
  {Kerber}}]{Girardi_etal09}
{Girardi} L., {Rubele} S., {Kerber} L., 2009, \mnras, 394, L74

\bibitem[{{Grevesse} \& {Sauval}(1998)}]{Grevesse_sauval98}
{Grevesse} N., {Sauval} A.~J., 1998, \ssr, 85, 161

\bibitem[{{Kamath}, {Karakas} \& {Wood}(2012){Kamath}, {Karakas}, \&
  {Wood}}]{Kamath_etal12}
{Kamath} D., {Karakas} A.~I., {Wood} P.~R., 2012, \apj, 746, 20

\bibitem[{{Karachentsev}, {Aparicio} \& {Makarova}(1999){Karachentsev},
  {Aparicio}, \& {Makarova}}]{Karachentsev_etal99}
{Karachentsev} I., {Aparicio} A., {Makarova} L., 1999, \aap, 352, 363

\bibitem[{{Lauterborn}, {Refsdal} \& {Weigert}(1971){Lauterborn}, {Refsdal}, \&
  {Weigert}}]{Lauterborn_etal71}
{Lauterborn} D., {Refsdal} S., {Weigert} A., 1971, \aap, 10, 97

\bibitem[{{Lee} \& {Kim}(2000)}]{Lee_etal00}
{Lee} M.~G., {Kim} S.~C., 2000, \aj, 119, 777

\bibitem[{{Marigo} {et~al}\mbox{.}(2008){Marigo}, {Girardi}, {Bressan},
  {Groenewegen}, {Silva}, \& {Granato}}]{Marigo_etal08}
{Marigo} P., {Girardi} L., {Bressan} A., {Groenewegen} M.~A.~T., {Silva} L.,
  {Granato} G.~L., 2008, \aap, 482, 883

\bibitem[{{Massevitch} {et~al}\mbox{.}(1979){Massevitch}, {Popova}, {Tutukov},
  \& {Iungelson}}]{Massevitch_etal79}
{Massevitch} A.~G., {Popova} E.~I., {Tutukov} A.~V., {Iungelson} L.~R., 1979,
  \apss, 62, 451

\bibitem[{{Meakin} \& {Arnett}(2007)}]{Meakin_arnett07}
{Meakin} C.~A., {Arnett} D., 2007, \apj, 667, 448

\bibitem[{{Mermilliod} {et~al}\mbox{.}(1992){Mermilliod}, {Rosvick},
  {Duquennoy}, \& {Mayor}}]{Mermilliod_etal92}
{Mermilliod} J.-C., {Rosvick} J.~M., {Duquennoy} A., {Mayor} M., 1992, \aap,
  265, 513

\bibitem[{{Momany} {et~al}\mbox{.}(2014){Momany}, {Clemens}, {Bedin},
  {Gullieuszik}, {Held}, {Saviane}, {Zaggia}, {Monaco}, {Montalto}, {Rich}, \&
  {Rizzi}}]{Momany_etal14}
{Momany} Y. {et~al.}, 2014, \aap, 572, A42

\bibitem[{{Momany} {et~al}\mbox{.}(2005){Momany}, {Held}, {Saviane}, {Bedin},
  {Gullieuszik}, {Clemens}, {Rizzi}, {Rich}, \& {Kuijken}}]{Momany_etal05}
---, 2005, \aap, 439, 111

\bibitem[{{Momany} {et~al}\mbox{.}(2002){Momany}, {Held}, {Saviane}, \&
  {Rizzi}}]{Momany_etal02}
{Momany} Y., {Held} E.~V., {Saviane} I., {Rizzi} L., 2002, \aap, 384, 393

\bibitem[{{Raghavan} {et~al}\mbox{.}(2010){Raghavan}, {McAlister}, {Henry},
  {Latham}, {Marcy}, {Mason}, {Gies}, {White}, \& {ten
  Brummelaar}}]{Raghavan_etal10}
{Raghavan} D. {et~al.}, 2010, \apjs, 190, 1

\bibitem[{{Robertson}(1972)}]{Robertson72}
{Robertson} J.~W., 1972, \apj, 173, 631

\bibitem[{{Sana}, {Gosset} \& {Evans}(2009){Sana}, {Gosset}, \&
  {Evans}}]{Sana_etal09}
{Sana} H., {Gosset} E., {Evans} C.~J., 2009, \mnras, 400, 1479

\bibitem[{{Saviane} {et~al}\mbox{.}(2002){Saviane}, {Rizzi}, {Held},
  {Bresolin}, \& {Momany}}]{Saviane_etal02}
{Saviane} I., {Rizzi} L., {Held} E.~V., {Bresolin} F., {Momany} Y., 2002, \aap,
  390, 59

\bibitem[{{Schlegel}, {Finkbeiner} \& {Davis}(1998){Schlegel}, {Finkbeiner}, \&
  {Davis}}]{Schlegel_etal98}
{Schlegel} D.~J., {Finkbeiner} D.~P., {Davis} M., 1998, \apj, 500, 525

\bibitem[{{Seaton}(1979)}]{Seaton79}
{Seaton} M.~J., 1979, \mnras, 187, 73P

\bibitem[{{Skillman}, {Terlevich} \& {Melnick}(1989){Skillman}, {Terlevich}, \&
  {Melnick}}]{skillman_etal89}
{Skillman} E.~D., {Terlevich} R., {Melnick} J., 1989, \mnras, 240, 563

\bibitem[{{Stothers} \& {Chin}(1991)}]{Stothers_etal91}
{Stothers} R.~B., {Chin} C.-W., 1991, \apj, 374, 288

\bibitem[{{Tang} {et~al}\mbox{.}(2014){Tang}, {Bressan}, {Rosenfield},
  {Slemer}, {Marigo}, {Girardi}, \& {Bianchi}}]{Tang_etal14}
{Tang} J., {Bressan} A., {Rosenfield} P., {Slemer} A., {Marigo} P., {Girardi}
  L., {Bianchi} L., 2014, \mnras, 445, 4287

\bibitem[{{Viallet} {et~al}\mbox{.}(2013){Viallet}, {Meakin}, {Arnett}, \&
  {Moc{\'a}k}}]{Viallet_etal13}
{Viallet} M., {Meakin} C., {Arnett} D., {Moc{\'a}k} M., 2013, \apj, 769, 1

\bibitem[{{Walmswell}, {Tout} \& {Eldridge}(2015){Walmswell}, {Tout}, \&
  {Eldridge}}]{Walmswell_etal15}
{Walmswell} J.~J., {Tout} C.~A., {Eldridge} J.~J., 2015, \mnras, 447, 2951

\end{thebibliography}
%
%to be uncommented before sending to editor
%\input{dualclump.bbl}
%
\label{lastpage}
\end{document}